\documentclass[a4paper,fleqn,usenatbib]{mnras}

\usepackage{newtxtext,newtxmath}

\usepackage[T1]{fontenc}
\usepackage{ae,aecompl}

\usepackage{graphicx}	
\usepackage{amsmath}	
\usepackage{amssymb}	
\usepackage{multirow}

\newcommand{\angstrom}{\mbox{\normalfont\AA}}

\title[Spatial Distribution of Star Formation in MaNGA]{SDSS-IV MaNGA: The Spatial Distribution of Star Formation and its Dependence on Mass, Structure and Environment}

\author[A. Spindler et al.]{
Ashley Spindler$^{1}$\thanks{E-mail: ashley.spindler@open.ac.uk},
David Wake$^{1,2}$,
Francesco Belfiore$^{3}$,
Matthew Bershady$^{4}$,
\newauthor
Kevin Bundy$^{3}$,
Niv Drory$^{5}$,
Karen Masters$^{6}$,
Daniel Thomas$^{6}$,
Kyle Westfall$^{3}$,
\newauthor
Vivienne Wild$^{7}$
\\
$^{1}$ The Open University, Walton Hall, Milton Keynes, MK7 6AA, UK.\\
$^{2}$ Department of Physics, University of North Carolina Asheville, One University Heights, Asheville, NC 28804, USA.\\
$^{3}$ University of California Observatories - Lick Observatory, University of California Santa Cruz, 1156 High St., Santa Cruz, CA 95064, USA.\\
$^{4}$ Department of Astronomy, University of Wisconsin, 475 N. Charter St., Madison, WI 53706, USA.\\
$^{5}$ McDonald Observatory, University of Texas at Austin, 1 University Station, Austin, TX 78712-0259, USA.\\
$^{6}$ Institute of Cosmology and Gravitation, University of Portsmouth, Portsmouth, UK.\\
$^{7}$ School of Physics and Astronomy, University of St Andrews, North Haugh, St Andrews, KY16 9SS, UK.\\
}

\date{Accepted XXX. Received YYY; in original form ZZZ}

\pubyear{2015}

\begin{document}
\label{firstpage}
\pagerange{\pageref{firstpage}--\pageref{lastpage}}
\maketitle

\begin{abstract}
We study the spatially resolved star formation of 1494 galaxies in the SDSSIV-MaNGA Survey. SFRs are calculated using a two-step process, using $H_\alpha$ in star forming regions and $D_n4000$ in regions identified as AGN/LI(N)ER or lineless. The roles of secular and environmental quenching processes are investigated by studying the dependence of the radial profiles of specific star formation rate on stellar mass, galaxy structure and environment. We report on the existence of `Centrally Suppressed' galaxies, which have suppressed SSFR in their cores compared to their disks. The profiles of centrally suppressed and unsuppressed galaxies are distibuted in a bimodal way. Galaxies with high stellar mass and core velocity dispersion are found to be much more likely to be centrally suppressed than low mass galaxies, and we show that this is related to morphology and the presence of AGN/LI(N)ER like emission. Centrally suppressed galaxies also display lower star formation at all radii compared to unsuppressed galaxies. The profiles of central and satellite galaxies are also compared, and we find that satellite galaxies experience lower specific star formation rates at all radii than central galaxies. This uniform suppression could be a signal of the stripping of hot halo gas in the process known as strangulation. We find that satellites are not more likely to be suppressed in their cores than centrals, indicating that the core suppression is an entirely internal process. We find no correlation between the local environment density and the profiles of star formation rate surface density.
\end{abstract}

\begin{keywords}
galaxies: star formation -- galaxies: evolution -- galaxies: structure -- galaxies: bulges -- galaxies: groups: general -- galaxies: clusters: general
\end{keywords}



\section{Introduction}

In the last two decades, large scale spectroscopic surveys (such as SDSS, \cite{2000AJ....120.1579Y}, GAMA, \cite{2011MNRAS.413..971D} and zCOSMOS, \cite{2007ApJS..172...70L}) have been a driving force in extragalactic astronomy. One of the principal results of these surveys is the characterisation of the bimodality in galaxy populations across a variety of galaxy properties. Morphological type, colour, star formation rate, stellar population age and gas content have all been shown to be strongly bimodal (\cite{2006MNRAS.373..469B, 2004ApJ...615L.101B, 2005ApJ...629..143B, 2003ApJ...592..819B, 2004ApJ...600..681B, 2009ARA&A..47..159B, 2010AJ....139.2097P}). Broadly, galaxies can be split into two groups; star forming galaxies which are typically low density, disk-like in shape and blue in colour, and quiescent galaxies, which are more compact than star forming galaxies, generally do not host spiral shapes and are red in colour. Quiescent galaxies also typically contain older stellar populations than star forming galaxies \citep{2005ApJ...621..673T, 2009ARA&A..47..159B}. \cite{2007ApJ...665..265F} found that while the number density of blue galaxies has remained constant since $z \sim 1$, the number density of red galaxies has increased. These observations suggest then that there are physical processes that move galaxies from the Star Forming type to the Quiescent type. In this work we explore the shut down of star formation, or 'quenching' in local galaxies. We explore processes that shut down star formation at the local and global scale, and which act on different time scales.

In recent years, a new generation of integral field spectroscopy (IFS) surveys have been employed to study the evolution of galaxies and by extension the process of quenching. These IFS surveys (such as CALIFA, \cite{2012A&A...538A...8S}, MaNGA, \cite{2015ApJ...798....7B},  and SAMI, \cite{2015MNRAS.447.2857B}) use monolithic or multi-object spectrographs, and fibre optic bundles (or integral field units, IFUs) to observe galaxies both spatially and spectrally. The resulting data cubes provide spatially resolved information about the spectral make-up of the galaxy, allowing astronomers to study the spatial distribution of galaxy properties such as star formation, metallicity, kinematics and stellar age.

It has been suggested for some time that there are multiple channels by which galaxies can quench. Broadly speaking, there has been some consensus in the literature to divide processes into two channels, those dependent on stellar mass and those that rely on environment \citep{1977ApJ...211..638S, 1977MNRAS.179..541R, 2010ApJ...721..193P, 2013MNRAS.429.2212M, 2014MNRAS.440..889S, 2015MNRAS.450..435S, 2016MNRAS.461.3111B, 2017MNRAS.466.2570B}. Mass-quenching refers to the mechanisms that shut down star formation due to the intrinsic properties of the galaxy, such as radio-mode feedback from AGN, morphological quenching, bar quenching and halo-shock heating \citep{2006MNRAS.370..645B, 2007MNRAS.382.1415S, 2011MNRAS.411.2026M, 2012ARA&A..50..455F, 2012Natur.485..213P, 2014ARA&A..52..589H, 2015A&A...580A.116G, 2016MNRAS.461.3111B, 2017MNRAS.466.2570B}. Environmental-quenching refers to the mechanisms related to the extrinsic properties of a galaxy, these include ram pressure stripping, tidal stripping, galaxy harassment and strangulation \citep{1972ApJ...176....1G, 1999MNRAS.308..947A, 2000ApJ...540..113B, 2002MNRAS.334..673L, 2008MNRAS.389.1619F, 2008MNRAS.383..593M, 2008MNRAS.387...79V, 2015A&A...576A.103B, 2015Natur.521..192P, 2017arXiv170508452G}.
	
Interestingly however, it has been shown by some authors that mass and environment quenching may in fact be part of the same mechanism. For example \cite{2015ApJ...800...24K} found that central galaxies in groups also respond to the environmental processes that are typically only associated with satellites, they go on to suggest that the differences in apparent mass dependences of satellite and central quenching occur because the properties that determine satellite quenching (e.g., dark matter halo mass, group centric distance, local overdensity) are independent of satellite stellar mass. \cite{2016ApJ...818..180C} and \cite{2017MNRAS.469.3670S} both suggest that environmental processes work in tandem with mass and morphological quenching mechanisms in driving the evolution of satellite galaxies in groups.

There are a number of physical processes which act on galaxies in dense environments, which have been widely studied in the literature. Ram pressure stripping refers to the removal of gas from a galaxy due to super sonic heating in the intracluster medium \citep{1972ApJ...176....1G, 1994AJ....107.1003C, 1982ARA&A..20..547F, 2000ApJ...541..542M,2001ApJ...548...97S, 1985AJ.....90.2445G,2011MNRAS.415.1797C}. Ram pressure stripping leads to a confinement of star formation to the centres of galaxies, as it predominantly acts on the outer disk of later type galaxies (\cite{2004ApJ...613..866K, 2004ApJ...613..851K, 2012A&A...544A.101C}). Similarly, galaxies may be subject to tidal harrassment from the surrounding dark matter halo and neighbouring galaxies, which affects star formation by removing gas from the disks or driving it into the galaxy bulges \citep{1989Natur.340..687H, 2015MNRAS.448.1107M}.

If a galaxies outer halo of gas is stripped away, it will lose the ability to replenish the gas it uses in star formation, causing an eventual shut down in star formation often referred to as starvation or strangulation \citep{1980ApJ...237..692L, 2008MNRAS.383..593M, 2015Natur.521..192P}. Interestingly, strangulation is predicted to have a different spatial pattern than gas stripping, occurring uniformly over the entire galaxy to produce anaemic spirals, as opposed to preferentially shutting down star formation in the disks or bulges of galaxies \citep{1991PASP..103..390V, 2002AJ....124..777E}.

The existence of mass-based and secular quenching has been widely established in the literature, but the understanding of the underlying physics on the other hand is not. \cite{2008ApJ...688..770F, 2012ApJ...753..167B, 2012ApJ...760..131C, 2012MNRAS.425..273P, 2012ApJ...751L..44W} and \cite{2014MNRAS.441..599B} all point out the strong link between the presence of a large bulge and the likelihood that a galaxy will be quenched. \cite{2009ApJ...707..250M} showed that the build up of a spheroidal components from mergers or other processes can stabilise the gas in a galaxy against collapse and fragmentation. This prevents star formation and causes early type galaxies to become red and dead. \cite{2015MNRAS.450..435S} found that quenching time-scales are correlated with galaxy morphology. Bars have also been linked to low the shut down of star formation in galaxies, both on a global scale and with the central few kpc of the galaxy core \citep{2011MNRAS.411.2026M, 2015A&A...580A.116G}

The large bulges in quenched galaxies leads to the assumption that supermassive black holes may play a role in quenching, as the black hole mass is well correlated with  bulge mass (\cite{2003ApJ...589L..21M, 2004ApJ...604L..89H, 2013ApJ...764..184M}). It has been show by \cite{2012ARA&A..50..455F} that radio-mode AGN are capable of inflating large bubbles of ionised gas, which could play an important role in regulating star formation and gas accretion. However, no link has been found between the presence of a radiative mode AGN and a suppression of star formation (\cite{2012MNRAS.425L..66M, 2014A&A...562A..21C, 2015A&A...580A.102C}).

It appears then, from the mechanisms that drive mass based and environment based quenching, that they should provide opposing signals in galaxies. So-called ``Inside-out" and ``outside-in" quenching has been discussed in the literature (\cite{2015Sci...348..314T, 2015ApJ...804..125L}). The environment channel may demonstrate an outside-in signal, whereby the cold gas is stripped from the outer disks or driven into the centre by tidal interactions, which would present enhanced star formation in the galaxy cores with respect to the outskirts. Mass quenching, if driven by AGN feedback or bulge growth, would instead demonstrate an inside-out quenching pattern, as the AGN quenches the star formation in the galaxy bulges first.

Thanks to the next generation integral field spectroscopy surveys we can now study the effects of quenching at spatially resolved scales and identify the signals for both the mass based and environment based quenching mechanisms. \cite{2017MNRAS.466.2570B} have already shown the presence of inside-out quenching with their study of ``central low ionisation emission region" (cLIER) galaxies, which they show could be green valley galaxies in the process of quenching. The outside-in process, instead, has been observed in MaNGA through stellar population analysis by \cite{2017MNRAS.466.4731G} who find slightly positive age gradients in early-type galaxies pointing towards outside-in progression of star formation. This pattern was found to be independent of environmental density in \cite{2017MNRAS.465..688G} and \cite{2017MNRAS.465.4572Z}. \cite{2017MNRAS.464..121S}, used the Sydney-AAO Multi-Object Integral Field Spectrograph (SAMI), to show that increasing local density correlated with reduced star formation in the outskirts of galaxies. Conversely, \cite{2013MNRAS.435.2903B} found no evidence of environmental quenching on a sample of galaxies studied using their $H_\alpha$ profiles, however this sample size was much smaller than \cite{2017MNRAS.464..121S} with only 18 galaxies in the former and 201 galaxies in the latter. Narrow band imaging of $H_\alpha$ has been used to study the environmental dependence of star formation in dense environments. In the Virgo cluster \cite{2004ApJ...613..866K} showed that approximately half of their sample of 84 galaxies had truncated star formation, and 10\% had star formation rates which were uniformly suppressed. In the Calar Alto Legacy Integral Field Area survey (CALIFA) \cite{2013ApJ...764L...1P} showed that massive galaxies grew their mass inside-out by using stellar population spectral sysnthesis to find spatially and time resolved star formation histories. \cite{2017arXiv170606119G} also studied spatially resolved star formation histories of a morphologically diverse sample of galaxies and found that galaxy formation happens very rapidly and in the past it was the central regions of early type galaxies where star formation was at its most intense. In addition, \cite{2017ApJ...838..105L} found evidence of bar induced star formation in the centres of so-called `turnover galaxies', which exhibit a rejuvenated stellar populations in their cores.

In this paper we use a large sample of 1368 Star Forming and Composite AGN/Star Forming galaxies from the Fourth Sloan Digital Sky Survey Mapping Nearby Galaxies at APO (SDSSIV-MaNGA, \cite{2015ApJ...798....7B, 2017AJ....154...28B}) survey to study the spatial distribution of star formation and its dependence on stellar mass, core velocity dispersion, morphology and environment. We calculate star formation rates using dust corrected $H_{\alpha}$ measurements and the $D_n4000$ spectral index and investigate the shapes of the galaxy's specific star formation rate profiles and investigate whether there is an inside-out or outside-in suppression of star formation with respect to galaxy's internal and external properties.

This work is complemented by a parallel paper (Belfiore et al., submitted), which studies the sSFR profiles in the Green Valley and in central LIER galaxies.

This work is structured as follows. In Section \ref{Data} we discuss the MaNGA survey and our sample selection criteria. In Section \ref{SFRs} we construct our star formation rates using dust corrected $H_\alpha$ and show our model for using $D_n4000$ in regions of the galaxies where $H_\alpha$ is unreliable. In Section \ref{Results} we show our results for the specific star formation rate profiles and their dependence on a variety of galaxy properties, then in Section \ref{CenSupp} we split the galaxy sample in galaxies which are centrally quenched or star forming. Finally we conclude in Section \ref{Conc} and discuss the roles of environment and mass based quenching in relation to this work. We make use of a standard $\Lambda$CDM cosmology with $\Omega_m = 0.3$, $\Omega_{\Lambda} = 0.7$ and $H_0 = 70 km^{-1} s^{-1} Mpc^{-1}$.

\begin{figure}
	\includegraphics[trim = 0mm 0mm 0mm 0mm, clip, width=1.0\columnwidth]{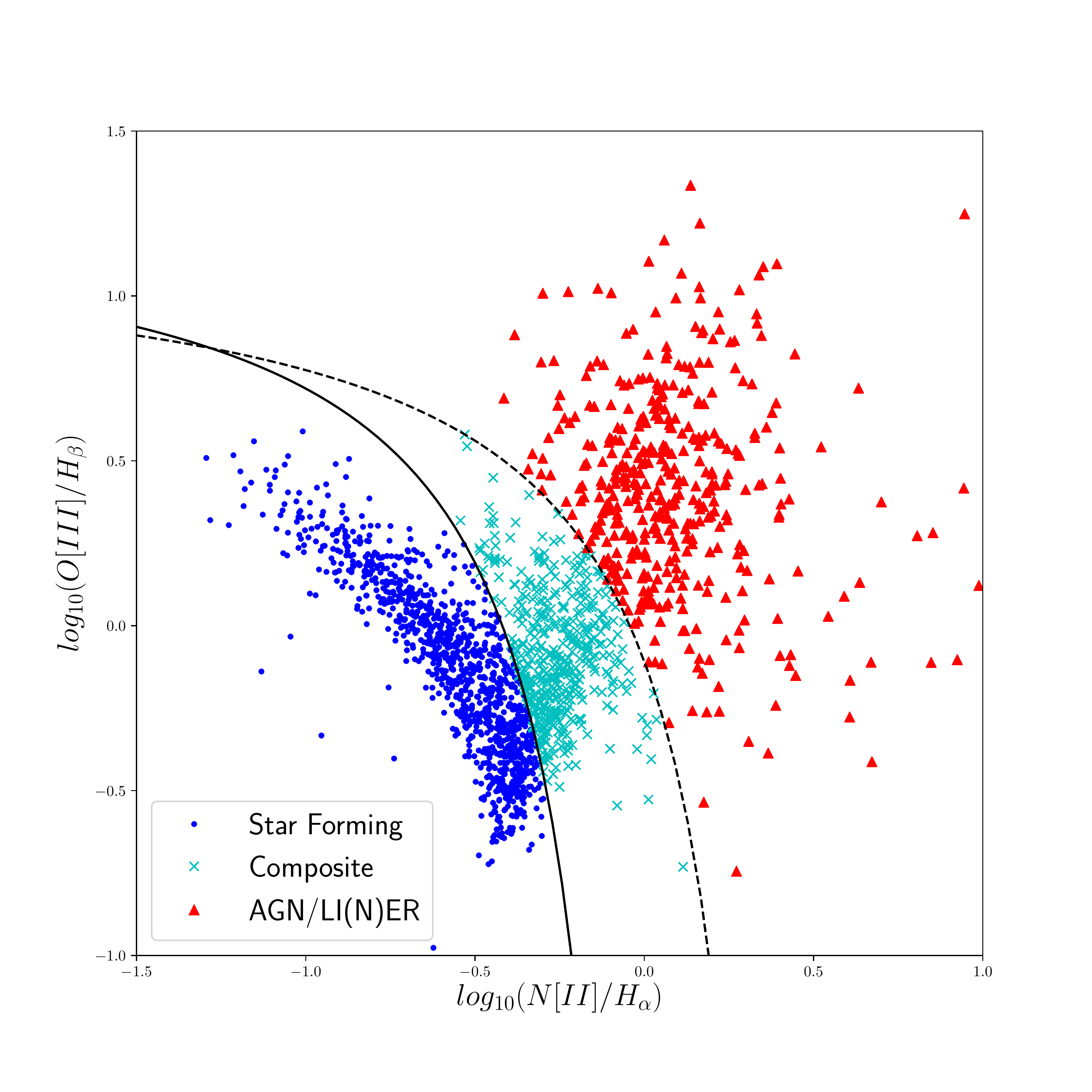}
	\centering
	\caption{\small
		The BPT Diagram for galaxies in the MaNGA survey. The positions of galaxies are calculated from the integrated flux over the entire IFU. Blue dots are the Star Forming Galaxies, cyan crosses and the composite galaxies and the red triangles are the AGN/LINER galaxies. The solid line is the relation from \protect\cite{2003MNRAS.346.1055K} and the dashed line is from \protect\cite{2001ApJ...556..121K}.
		\label{fig:BPT_Diagram}}
\end{figure}

\section{Data}
\label{Data}

\subsection{MaNGA Data}

Mapping Nearby Galaxies at APO \citep[MaNGA, ]{2015ApJ...798....7B, 2015AJ....150...19L, 2016AJ....152..197Y} is a multi-object IFU survey, one of the three projects under way as part of SDSS-IV \citep{2017AJ....154...28B} using the 2.5-meter Sloan Foundation Telescope at the Apache Point Observatory \citep[APO, ]{2006AJ....131.2332G}). The goal of MaNGA is to observe $\sim 10,000$ galaxies using a range of IFU bundle sizes \citep{2015AJ....149...77D}. Observations began in 2014 and will conclude in 2020. The galaxy sample is chosen to include galaxies with $M_* > 10^9 M_\odot$ and have a flat number density distribution as a function of mass, while having no cuts in morphology, colour or environment. MaNGA has three main subsamples, the Primary, Secondary and Colour-Enhanced samples. The Primary sample makes up 50\% of the target catalog, has a flat distribution in K-corrected i-band magnitude and has a spatial coverage of $1.5 r_e$ within the IFUs. The Secondary sample contains 33\% of the MaNGA sample, also has a flat distribution in $M_i$ but instead selects IFUs which cover galaxies out to 2.5 $r_e$. Finally, the Colour-Enhanced sample makes up the remaining 17\% of target galaxies, and is selected to sample galaxies from regions in the $NUV - i$ versus $M_i$ plane which are under sampled by the primary sample such as low-mass red galaxies and high-mass blue galaxies.

We study galaxies from Data Release 14 (DR14). Using a range of IFU sizes most of the galaxies have full spectral coverage up to $1.5$ half-light radii ($r_e$), though a subset are observed out to $2.5 r_e$. The IFU fibres are fed into the BOSS spectrograph, which has continuous coverage between $3600 \angstrom$ and $10300 \angstrom$, with a spectral resolution of $R \sim 2000$ \citep{2013AJ....146...32S, 2015AJ....149...77D}. The MaNGA observations are reduced into data cubes by the Data Reduction Pipeline (DRP, \cite{2016AJ....152...83L}) and then analysed using the Data Analysis Pipeline (DAP, Westfall et al. in prep). The DAP fits the continuum, emission lines, kinematics and spectral indices from the DRP data cubes. Throughout this paper we use the galaxy weights from \cite{2017arXiv170702989W}, which are used to correct the sample from magnitude limited to volume limited.

We make use of three of the products from the Data Analysis Pipeline (DAP, Westfall et al. in prep), the ALL binned data which combines the flux from all the spaxels in the data cube for maximum signal to noise, the VOR10 data which bins the spaxels into SNR>10 Voronoi bins and the NONE binned data which includes all of the spaxels in the data cubes individually. The ALL binned data is used when calculating our data cuts described in Section \ref{Sample}. We use the Voronoi binned data to calibrate our $D_n4000$-SSFR model and the unbinned data is used in the final analysis. In addition we have rerun the DAP to produce a additional map of each galaxy which contains a single spatial bin out to $0.125 r_e$, which is used to find the core velocity dispersion, $\sigma_0$, to match the definition used in \cite{2017MNRAS.468..333S}.

\subsection{Sample Selection}
\label{Sample}

DR14 contains 2791 galaxies across the primary, secondary, colour enhanced and ancillary samples. In this work we begin with the full MaNGA sample, with galaxies from the Primary, Secondary and Colour-Enhanced Samples.

We remove IFUs which contain two or more galaxies from the sample, which were identified by eye in the SDSS g-r-i imaging of the MaNGA galaxies, which cuts 153 fibre bundles from the sample. We do this to eliminate the need to calculate centres for both galaxies in order to find individual SFR profiles.

Throughout this work we wish to study galaxies which are dominated by different forms of ionising radiation, such as from star formation, Active Galactic Nuclei (AGN) and Low-Ionization (Nuclear) Emission Regions (LI(N)ER), or galaxies which are a composite of these emission types. As such, we measure the line intensities of $H_\alpha$, $H_\beta$, [$NII$] ($6585 nm$) and [$OIII$] ($5008 nm$) in the integrated fluxes of the DR14 data cubes and calculate the positions of these galaxies on the Baldwin-Phillips-Terlevich (BPT, \cite{1981PASP...93....5B}) diagram. We require that the emission line SNR in each of these lines be $> 2$ to accurately calculate their positions on the BPT diagram, the limiting factors in the signal-to-noise are the strengths of the $H_\beta$ and [$OIII$] lines. We divide the galaxies into five groups: Star Forming for galaxies which fall below the \cite{2003MNRAS.346.1055K} line, Composite for galaxies between the Kauffmann and \cite{2001ApJ...556..121K} lines, AGN/LI(N)ER for those above the Kewley line, Low SNR AGN for galaxies with low SNR in the $H_\beta$ and[$OIII$] lines but with integrated $SNR>3$ in $H_\alpha$ and [$NII$] with $\log_{10}(H_\alpha/NII) > 0.47$ and finally Lineless galaxies for those galaxies with low SNR in all four diagnostic lines. We find  1049 Star Forming galaxies and 435 Composite galaxies which we examine in the main bulk of this paper, in addition there are 428 AGN/LI(N)ER and 22 low SNR AGN galaxies which we study in Section \ref{AGNfeed}, and 719 Lineless galaxies which we discard from the sample. The BPT diagram for the DR14 sample is shown in Figure \ref{fig:BPT_Diagram} and shows the separations used in this sample selection. Finally, we remove from the sample galaxies which have total Specific Star Formation Rates (calculated using the model described in Section \ref{SFRs}) of $\log_{10}(SSFR)<-11.5$.

The above classification are different to Belfiore et al. (submitted), in which we use a spatially resolved BPT classifications. While the above work is interested in the roles of cLIER galaxies and their transition through the green valley, in this work we are interested in the much broader trends across the entire population. In this case we find that using the integrated flux to calculate the BPT class suits our needs, especially with the inclusion of the composite class which includes galaxies with star forming disks and AGN/LI(N)ER central regions which may be confused with only a SF-AGN/LI(N)ER cut. An alternative classification system in which we measured the BPT classification in the central 3" of each galaxy was tested, however we found that the majority of the galaxies which have different classes in this system were AGN/LI(N)ERs and lineless galaxies which are otherwise already removed from the sample due to low SSFRs.

A final cut is applied to the sample based on galaxy axis ratio. Edge-on disks with a $b/a<0.3$ are removed from the sample, as we have found that their radial profiles are poorly resolved. A total of 128 galaxies are removed based on this cut. The final sample is then composed of 1494 galaxies, 1016 of which are star forming, 364 are composite and 114 are AGN/LI(N)ER.

In addition to the core MaNGA data products we make use of the SDSS-MaNGA-Pipe3D (Pipe3D, \cite{2016RMxAA..52..171S, 2016RMxAA..52...21S}) value added catalog. The Pipe3D data products were developed using the pipeline described in \cite{2016RMxAA..52..171S} and  \cite{2016RMxAA..52...21S} and applied to DR14. We use the Single Stellar Population (SSP) cubes, which provide stellar mass surface density ($\log_{10}(M_\odot) arcsec-2$) maps of the galaxies in DR14.

\subsection{Other Catalogs}
\label{OtherCats}

We make use of two additional catalogs in the analysis of this work, the Yang Group Catalog (\cite{2007ApJ...671..153Y, 2008ApJ...676..248Y, 2009ApJ...695..900Y, 2012ApJ...752...41Y}) and the \cite{2006MNRAS.373..469B} Environment Density catalog.

The Yang Group Catalog uses a friends of friends algorithm to generate galaxy groups and clusters using SDSS DR7. Galaxies are matched into tentative groups and properties such as dark matter halo mass and group luminosity are calculated, from these properties the halo groups are recalculated to include nearby galaxies that fall within the halos. This iterative process continues until no new galaxies are added to groups. From this catalog we use the Central and Satellite galaxy classifications, the dark matter halo masses and the group luminosities. The galaxy classifications and halo masses are based on rankings of the galaxies luminosities.

There are a small number of galaxies in the MaNGA sample that are not in the SDSS DR7 (their NSA redshifts come from other sources) and so are not included in the Yang et al. catalog. We assign these galaxies central/satellite designations and group luminosities and halo masses by associating them with Yang et al groups where possible. If a non-DR7 MaNGA galaxy has a projected separation within $r_{180}$ of a group centre and a velocity within $\pm$ 1.5 times the group velocity dispersion then we associate it with the group. If there is no matching group then the galaxy becomes its own group. The galaxy is then designated as either the group central or a group satellite depending on whether or not it's r-band luminosity is the largest in the group. We then recalculate the group luminosity including the new galaxy and calculate the other group properties following Yang et al. prescription.

Finally, we make use of the environment densities around galaxies calculated in \cite{2006MNRAS.373..469B}. These densities are based on the distances to the 4th and 5th nearest neighbour galaxies with $M_r<-20 (h=0.7)$. The density is calculated as $\log_{10}(\Sigma) = 0.5*\log_{10}(\Sigma_4) + 0.5*\log_{10}(\Sigma_5)$, where $\Sigma_N=N/(pi*d_N^2)$ and $d_N$ is the distance to the Nth nearest neighbour. An important note here is that the matching between this catalog and the MaNGA data is not perfect, mainly owing to the redshift limits in the \cite{2006MNRAS.373..469B} galaxies. \cite{2006MNRAS.373..469B} is limited to $0.01 < z < 0.085$, which results in $15\%$ of our MaNGA sample not being assigned environment densities. Due to the relationship between stellar mass and redshift in MaNGA \citep{2017arXiv170702989W}, this means the galaxies without densities are mainly at higher masses.

\begin{figure}
	\includegraphics[trim = 0mm 0mm 0mm 0mm, clip, width=1.0\columnwidth]{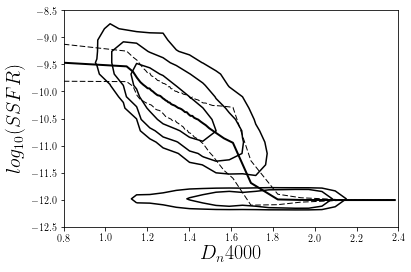}
	\centering
	\caption{\small
		Contours of the distribution of $D_n4000$ and SSFR, the contours represent the $1-$, $2-$ and $3-\sigma$ levels. The thick solid line is the mean fitted to the data we use for spaxels which are marked as composite or AGN/LINER from the BPT Diagram. Spaxels which we classify as low SNR are included in this model with an upper limit of $\log_{10}(SSFR) = -11.5$. The dashed lines are the standard deviation from the mean.
		\label{fig:D4000_Model}}
\end{figure}

\begin{figure}
	\includegraphics[trim = 0mm 0mm 0mm 0mm, clip, width=1.0\columnwidth]{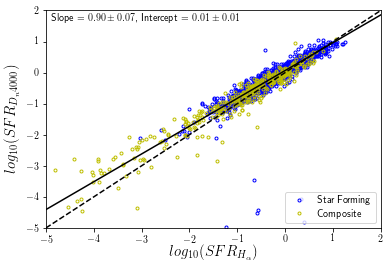}
	\centering
	\caption{\small
		We show the star formation rates calculated using just the H$_{\alpha}$ method and just the $D_n4000$ method for star forming and composite galaxies in MaNGA. The dashed line shows the 1-to-1 relation and the solid line shows the linear regression fit. We provide the slope and intercept of the fit in the top left corner, with errors calculated from 1000 bootstrap resamplings of the data.
		\label{fig:HaVD4000}}
\end{figure}

\begin{figure}
	\includegraphics[trim = 0mm 0mm 0mm 0mm, clip, width=1.0\columnwidth]{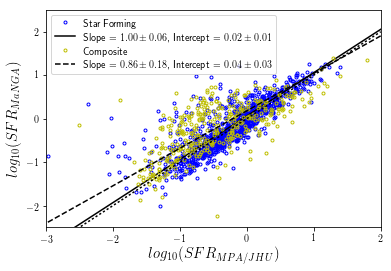}
	\centering
	\caption{\small
		Values of the star formation rates calculated using the method described here for Star Forming (blue) and Composite (yellow) MaNGA galaxies, compared with their star formation rates calculated in Brinchmann 04 for the MPA/JHU catalog. The dotted line shows the one-to-one relations, the solid line is the linear fit to the star forming galaxies and the dashed line is the fit to the composite galaxies. The parameters of the fits are show in the top left corner, with errors calculated from 1000 bootstrap resamplings.
		\label{fig:MangaVMPA_SFR}}
\end{figure}

\begin{figure*}
	\includegraphics[trim = 0mm 0mm 0mm 0mm, clip, width=1.0\textwidth]{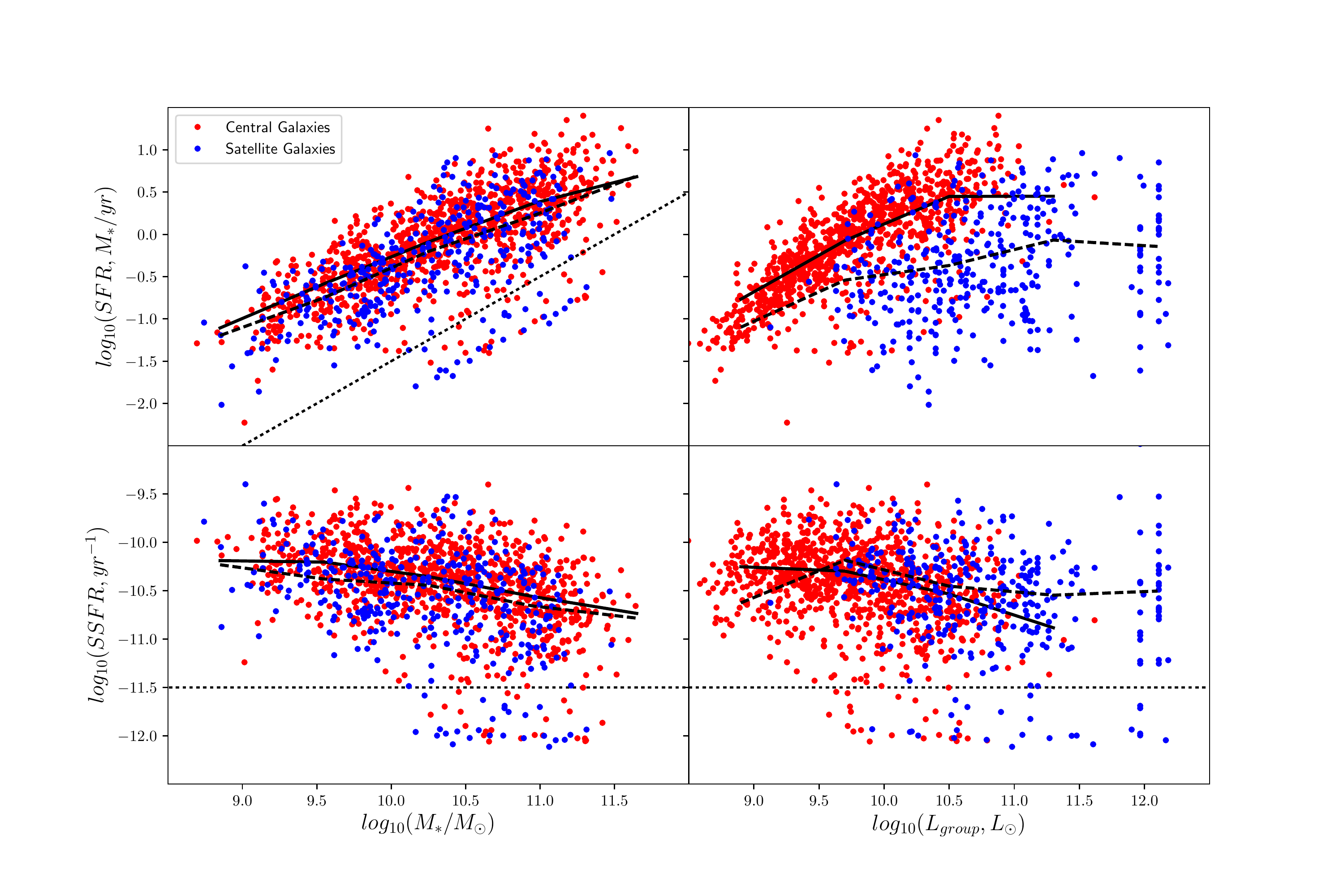}
	\centering
	\caption{\small
		We show the relationships between stellar mass in the left column, group luminosity in the right column, star formation rate in the top row and specific star formation rate in the bottom row, for galaxies with Star forming and Composite BPT types. Galaxies are coloured based on their environment, with centrals in red and satellites in blue. We include the mean values of SFR and SSFR at fixed $M_*$ and $L_{group}$ as solid lines for centrals and dashed lines for satellites. The dotted lines indicate the position of the sample cut in specific star formation rate at $\log_{10}(SSFR) = -11.5$.
		\label{fig:Mass_GroupL_SFR}}
\end{figure*}

\section{Star Formation Rates}
\label{SFRs}

In this Section we will present our method for producing spatially resolved maps of star formation. We use a two-source model, which calculates star formation rate from $H_\alpha$ emission in the first instance in spaxels which are classified as star forming in the BPT diagram. These SFRs are used to model the dependence of specific star formation rate on the strength of the 4000 {\AA} break ($D_n4000$). We then use this model to find the SFRs in spaxels with AGN and LINER contamination, and spaxels which are lineless, which would otherwise be missed in a model which relies only on $H_\alpha$ emission. We use star forming spaxels from both star forming and composite galaxies to ensure the SSFR-$D_n4000$ is as representative of our sample as possible. This method is inspired by the work of \cite{2004MNRAS.351.1151B} (B04) in the star formation estimations in the MPA/JHU DR7 catalog and allows us to include more galaxies than previous spatially resolved studies of star formation and study the star forming properties of galaxy bulges which would otherwise be removed due to contamination.

The final model will be applied to the DAP maps with no spatial binning, however it is important to begin with high signal-to-noise data so that we can detect very low levels of $H_\alpha$ emission and therefore allow our $D_n4000$-SSFR model to go to as low SSFRs as possible. As such we will begin our analysis using the Voronoi binned DAP products, which bins the spaxels into spatial regions which have a total r-band signal-to-noise ratio per bin $>10$. We apply an additional cut to this data and only use bins with SNR > 20.

Following from our previous BPT-classifications and the work of \cite{2016MNRAS.461.3111B}, we produce spatially resolved BPT diagnostic maps from the Voronoi binned data and unbinned data. Bins and spaxels are placed into 4 categories: Star Forming if they lie below the Kauffmann line, AGN/LI(N)ER if they lie above the Kauffmann line, lineless if they have $SNR < 2$ in the $H_\alpha$ or NII lines and low SNR AGN if the SNR for $H_\beta$ or $OIII$ is $< 2$, the SNR for $H_\alpha$ and $NII$ is $> 3$, and $\log_{10}(NII/H_\alpha) > 0.47$.

The star forming bins from the Voronoi maps have their star formation estimated using H$\rm{\alpha}$, as detailed in Section \ref{Halpha}, we then produce the model detailed in Section \ref{d4000} using these SFRs. The unbinned maps are then treated in the same way, with star forming spaxels using dust corrected $H_{\alpha}$ to estimate their SFRs and the AGN/LI(N)ER, low SNR and lineless spaxels estimated using the $D_n4000$ model.

\subsection{H$\rm{\alpha}$ SFRs}
\label{Halpha}

The $H_\alpha$ flux relates the emission from excited hydrogen clouds to the presence of high mass OB type stars, which dominate the light emitted in young stellar populations. $H_\alpha$ flux is readily absorbed and reprocessed by dust in the interstellar medium, we correct for this absorption by assuming a foreground dust screen and using the \cite{1989ApJ...345..245C} extinction law:

\begin{equation}
\label{dustcorr}
L_{H_\alpha} (\rm{Corrected}) = L_{H_\alpha} ((L_{H_\alpha}/L_{H_\beta})/2.8)^{2.36}
\end{equation}

This correction assumes a case B recombination at $T \sim 10,000$K and corrects the deviation from the theoretical ratio between the $H_\alpha$ and $H_\beta$ flux. The corrected $H_\alpha$ flux is converted into a SFR using the relation from \cite{1998ARA&A..36..189K}, for a \cite{1955ApJ...121..161S} IMF:

\begin{equation}
SFR(L_{H_\alpha}) = L_{H_\alpha} / 10^{41.1}
\end{equation}

\begin{figure*}
	\includegraphics[trim = 0mm 0mm 0mm 0mm, clip, width=1.0\textwidth]{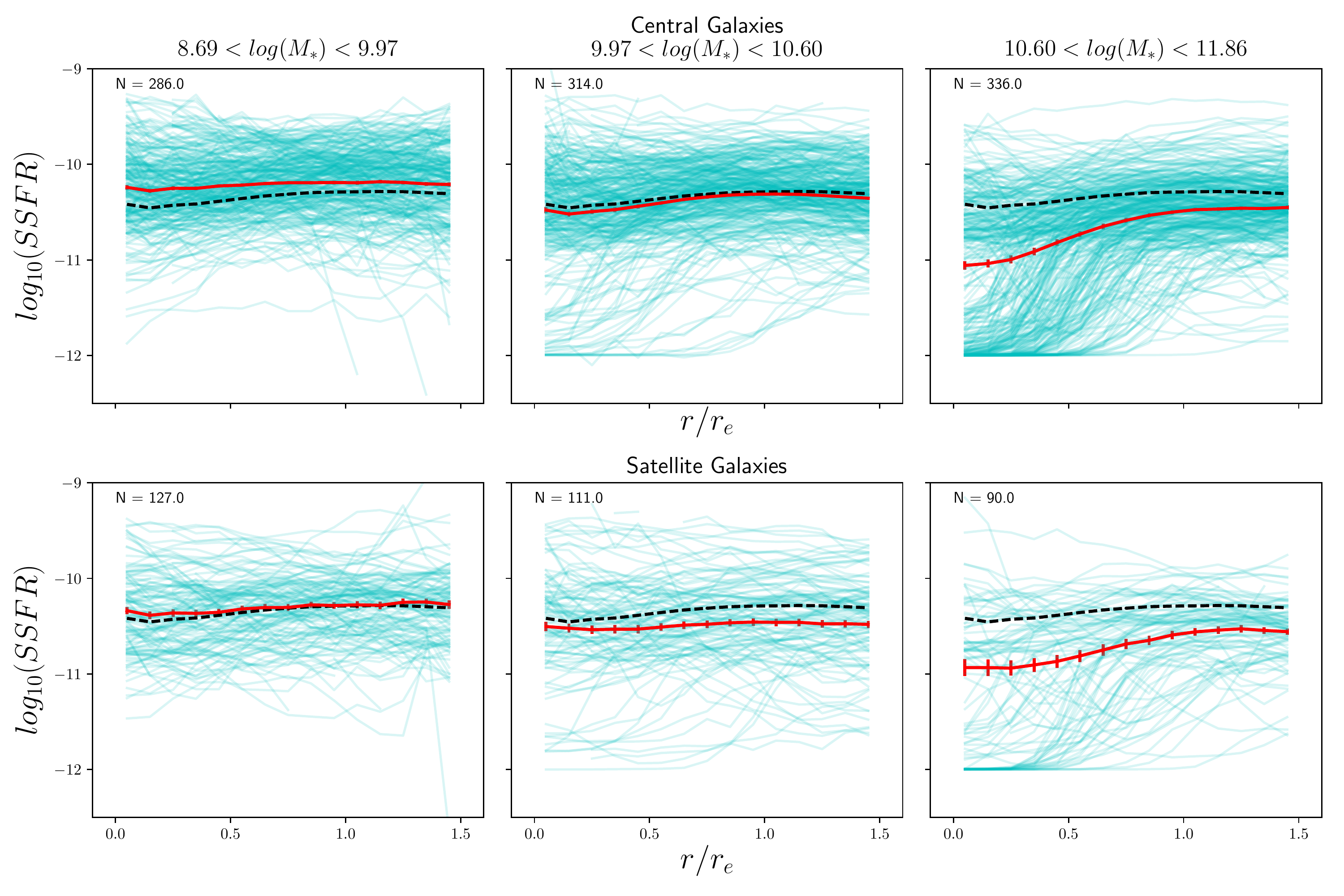}
	\centering
	\caption{\small
		The radial SSFR profiles in three bins of stellar mass. The individual profiles are shown by the cyan lines and the mean profile in the bin is shown by the solid red line. The dashed black line shows the mean profile of all galaxies in the sample. The number of galaxies in each bin is shown in the top left corner of each panel. The top row is the central galaxies and the bottom row is the satellite galaxies. The error bars are calculated from the scatter in 1000 bootstrap resamplings.
		\label{fig:Mass_Profiles}}
\end{figure*}

\begin{figure*}
	\includegraphics[trim = 0mm 0mm 0mm 0mm, clip, width=0.8\textwidth]{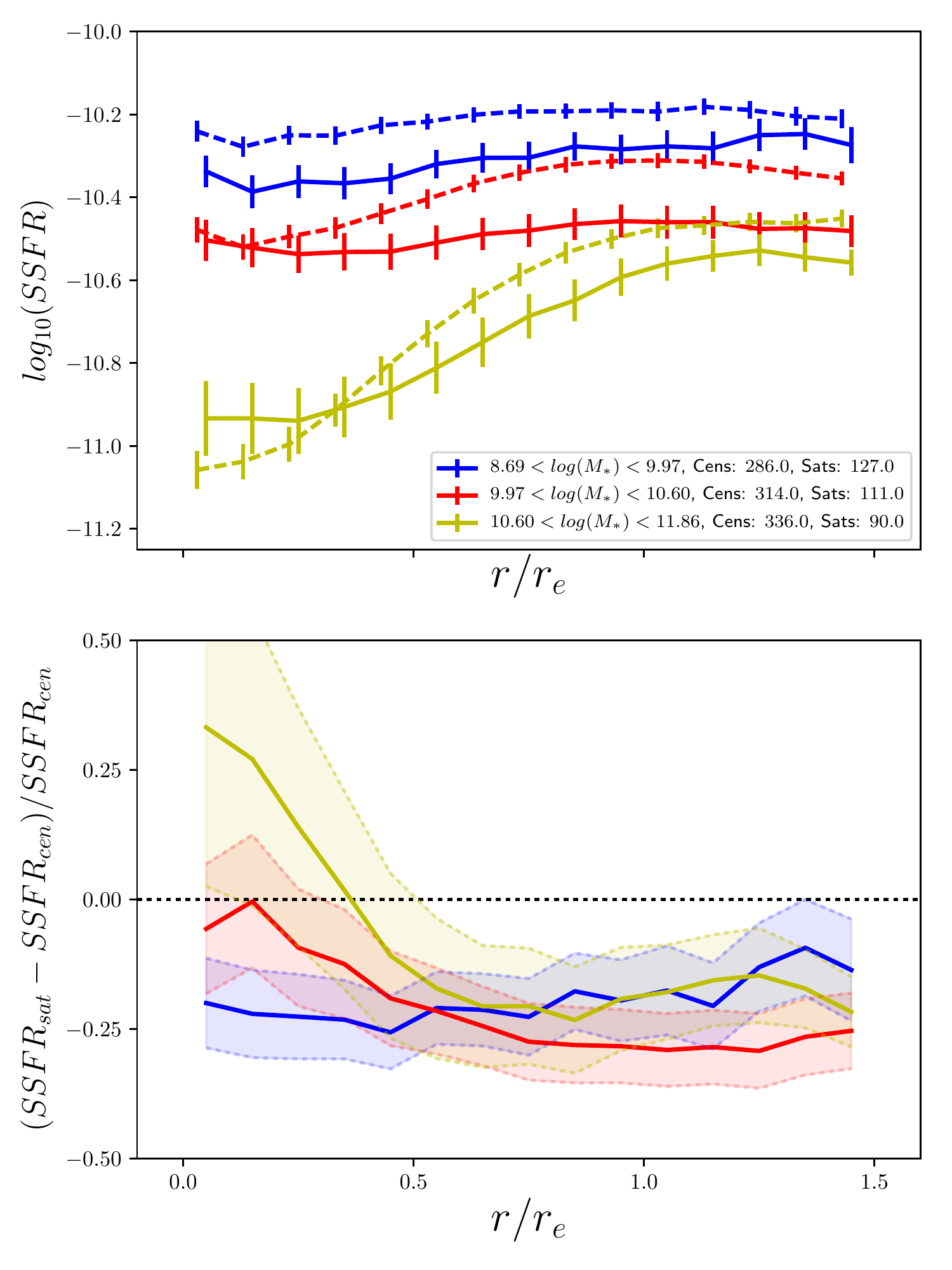}
	\centering
	\caption{\small
		(Top) The mean radial SSFR profiles of central (dashed) and satellite (solid) lines in bins of stellar mass. (Bottom) The fractional difference between the central and satellite mean profiles in bins of stellar mass. The shaded regions and error bars represent the $1-\sigma$ scatter in 1000 bootstrap resamplings.
		\label{fig:Difference_Mass}}
\end{figure*}

\subsection{$D_n4000$ SFRs}
\label{d4000}

In areas of the galaxy where there is contamination in the the $H_\alpha$ emission from AGN, LI(N)ER, old stellar populations and shocked gas we need a different estimator of Star Formation Rate. We also cannot simply ignore these portions of the galaxies, as the excess emissions often take place in important structures such as the bulge or bar. B04 showed that there is a relation between SSFR and $D_n4000$, which was used to estimate the SFRs of galaxies in DR4 and later DR7 of the Sloan Digital Sky Survey.

Using the Voronoi binned data we calculate the specific star formation rates using $H_\alpha$, in the regions which are diagnosed as star forming by the BPT diagram. As the star forming bins only cover a range of $D_n4000$ values ranging from 0.8 to 1.6 we also include the values of bins designated lineless, with a fixed upper limit SSFR of $\log_{10}(SSFR) = -12$. We require that the bins used here have a $SNR>20$, to ensure the quality of the model and to allow us to go to low values of $H_\alpha$. This approach is different from the one taken in Belfiore at al. (submitted), where radial annuli containing no spectroscopically-classified star forming regions are discarded in computing radial profiles. This difference should be taken into account where directly comparing the radial sSFR profiles of these two works.

In Figure \ref{fig:D4000_Model} we show the $D_n4000$-SSFR relation, the contours show the distribution of $D_n4000$ and the $H_\alpha$ predicted SSFRs in the star forming bins, the solid line shows the mean SSFR at fixed $D_n4000$ and the dashed lines are the first standard deviation from the mean. For galaxy regions which are marked as non-star forming, we assign a specific star formation rate by interpolating the $D_n4000$ measurement with the mean values from Figure \ref{fig:D4000_Model}. The SSFR decreases with increasing $D_n4000$ and flattens out at high values once it reaches the regime dominated by the lineless galaxies with high $D_n4000$ values. This flattening is artificial however, and is caused by the upper limit SSFR assigned to the lineless spaxels.

The value of the fixed SSFR limit applied at high $D_n4000$ values plays an important role in this work, as galaxies with old stellar populations will be assigned this value. At a qualitative level, we treat this limit as zero star formation, galaxies with this SSFR at certain points are treated as simply not forming stars whatsoever in those spaxels or radial bins. Quantitatively however, there is some dependence on the value of the limit on our work. For example, setting this value lower to $\log_{10}(SSFR) = -13$ has the effect of lowering total SSFRs of galaxies with $-11.5 < \log_{10}(SSFR) < -10.5$ by 0.14 dex on average, in addition to exaggerating the effects of any localised suppression of star formation within individual galaxies. However, we have tested using different values for the fixed SSFR limit and found that it has no effect on the conclusions of this paper.

To test the validity of this model, we compare the total SFRs predicted in the star forming spaxels in each galaxy using $H_\alpha$ and $D_n4000$ in Figure \ref{fig:HaVD4000}, with star forming galaxies in blue and composite galaxies in yellow. The two values of SFR agree very well, with most galaxies falling near the one to one relation with a scatter of 0.2 dex. Below $\log_{10}(SFR_{H_\alpha}) = -2$ the agreement is not 1-to-1, however these galaxies all have a very small number of spaxels ($< 10$) with both $H_\alpha$ and $D_n4000$ and so this can likely be attributed to the scatter in the $D_n4000$ model. We perform an orthogonal distance regression to fit a linear relation between the two values of star formation and find a very close to 1-to-1 fit, with a slope of $0.91 \pm 0.08$.

We compare the star formation rate in the MaNGA IFUs with the aperture corrected SFRs found in B04 for the MPA/JHU catalog in Figure \ref{fig:MangaVMPA_SFR}. The B04 total star formation rates are based using the broad band light from SDSS photometry to correct the single fibre measurement to a global value. The scatter from the one-to-one line is fairly tight, with a standard deviation of 0.35 dex. We provide two linear orthogonal distance regression fits to this comparison, one fit to the star forming galaxies and one to the composite galaxies. The star forming galaxies are fit very well, with a slope of $1.00 \pm 0.06$, we find that galaxies with lower star formation in the MPA/JHU are generally given higher SFR in our work, this most likely due to the use of the aperture correction to the 3" fibres in SDSS missing star formation which is present the MaNGA IFUs. For the composite galaxies we find the linear fit is worse than SF galaxies, but still close to 1-to-1 with a slope of $0.86 \pm 0.18$ and a scatter of 0.5 dex. As we will show in Section \ref{AGNfeed}, composite galaxies are more likely to have suppressed star formations in their centres but still be forming stars in their disks, as the MPA/JHU values are based on the fibre readings at the centre of the galaxies they would not pick up the extra star formation in the galaxy disk.

Throughout the rest of this paper we use the combination of $H_\alpha$ and $D_n4000$ star formation rates for our analysis. We note that when the analysis is performed using just the $D_n4000$ predictions for star formation there is no qualitative difference on the conclusions presented here.

\section{Results}
\label{Results}

\subsection{Global Properties}

We begin by studying the global properties of galaxies in MaNGA. We calculate the integrated SFR, SSFR and Stellar Masses of star forming and composite galaxies from the IFUs using the ALL binned DAP MAPs, and plot their relationships along with their group luminosities from the Yang Catalogue in Figure \ref{fig:Mass_GroupL_SFR}. We plot central galaxies from Yang in red and satellites in blue and show the mean relations for those galaxies in each panel with solid and dashed lines, respectively. We include galaxies which fall below our sample cut in SSFR, which is shown by the straight dashed line in the top left and bottom panels.

In the top left panel of Figure \ref{fig:Mass_GroupL_SFR} we show the ${M_*}$-SFR relation. We can clearly see the so-called 'Main Sequence of Star Formation' is present in this plot, as well as galaxies which fall into the 'green valley' (the region just above and below the SSFR cut). Below the SSFR cut we see galaxies with upper limit SFR which would make up the 'red sequence' of quiescent galaxies, however as these are upper limits it is important to note that this region of the plot would appear more cloud like with accurate estimates of star formation. The mean SFRs of the centrals and satellites are shown, with the satellites having lower SFR at fixed mass than the centrals, with an overall difference in the means of $0.1 \pm 0.03$ dex. These results are echoed in the bottom left panel, which shows the ${M_*}$-SSFR relation, with a difference in the means of $0.09 \pm 0.02$ dex. We again see that the satellites have lower SSFR than the central galaxies. There is a downward trend in the SSFR at fixed mass for both centrals and satellite galaxies.

In the top and bottom right panels of Figure \ref{fig:Mass_GroupL_SFR} we show the relationships of group luminosity with SFR and SSFR. For central galaxies these relationships are broadly similar to those with mass, as the luminosity of a group is tightly correlated with stellar mass for all but the most luminous groups. The satellite galaxies however are much more spread out in the $L_{group}$-SFR plane, as low mass satellites with low SFR can reside in very luminous groups, compared to centrals.

More massive star forming galaxies have lower specific star formation rates that low mass star forming galaxies, as seen in Figure \ref{fig:Mass_GroupL_SFR}, and quenched galaxies are also typically found at higher masses. This begs the question, what processes are taking place within more massive galaxies that are shutting down star formation. In the next sections we will study the mean radial profiles of specific star formation rates to investigate the mechanisms of star formation shut down, particularly whether the shut-down is inside-out or outside-in.

\subsection{SSFR Profiles at fixed $M_*$}
\label{Mass}

We wish to study the effects of internal and external processes on the distribution of star formation in galaxies within our sample. To test the effect of internal processes, we will investigate the mean profiles of galaxies in bins of stellar mass, core velocity dispersion and S\'ersic index, and to test for external environmental effects we will compare central and satellite galaxies. To investigate the distribution of star formation we choose to study the radial profiles of the specific star formation rates between $0-1.5 r_e$. For each galaxy we separate the star formation maps calculated in Section \ref{SFRs} into 15 bins of elliptical radius, each $0.1 r_e$ in width, from the centre of the galaxy. We calculate the mean SSFR of all the spaxels in each radius bin to find the radial profile of each galaxy.

An alternative way to calculate the radial profiles would be to integrate the light into elliptical radial bins, which can be done when processing datacubes with the DAP. We have tested this and found that it does not change the conclusions of this paper, so we choose to use the method described above.

We choose to calculate our radial profiles out to $1.5 r_e$ to ensure that the profiles are complete for each galaxy. While it is possible to extend these profiles out beyond this point, particularly for galaxies in the Secondary MaNGA sample which are assigned an IFU to cover out to $2.5 r_e$ and for edge on spirals which have radii going out to $5-6 r_e$, the vast majority of galaxies do not have the signal-to-noise at these larger radii to calculate a reliable star formation rate. We find that given our signal-to-noise cuts on the emission lines and $D_n4000$ that 80\% of galaxies are covered out to $1.5 r_e$ and this number falls to 50\% at $2.0 r_e$. Galaxies which are covered out to these larger radii tended to be assigned one of the larger IFUs and are preferentially from the Secondary galaxy sample, they are also typically more edge on disks.

In Figure \ref{fig:Mass_Profiles} we plot the radial SSFR profiles of central and satellite galaxies, in bins of stellar mass. The bins are chosen such that the total number of galaxies between centrals and satellites in each bin is constant. We show the individual profiles from $0-1.5 r_e$ in cyan, the mean profile of each bin in red, with errors calculated from 1000 bootstrap resamplings, and the mean profile of all galaxies in the sample as a black dashed line in each panel to guide the eye and provide a point of reference.

In the lowest mass bin we see that the central and satellite profiles are largely flat, and while there are individual profiles that rise or fall with increasing radius the mean profiles remain constant. In the medium mass bin the mean profile is still rather flat, but we see that the central mean profile has been pulled down slightly by a population of galaxies which have low central SSFRs, while the satellites remain flat. The differences in the centres of galaxies are subtle, and we explore this effect further in Section \ref{EnvQuench}. In the highest mass bin the galaxies with suppressed cores have significantly altered the shape of the mean profiles, which now exhibits a two-component shape with low SSFR in the centre and a flat profile outside of 1 $r_e$.

We can see by comparing the mean profiles in each bin with the full sample mean that the total specific star formation rate drops as stellar mass increases and that the galaxies which have suppressed star formation in their cores are mostly isolated to high masses. Figure \ref{fig:Mass_Profiles} also displays a bimodality, particularly at high masses,  between two galaxy classes, those with relatively flat profiles and those which have suppressed star formation in their centres. However there is a difference regarding the extent of the suppression from the centre of the galaxy, with some galaxies beginning to show suppression at very small radii, and others at more intermediate radii.

\begin{figure}
	\includegraphics[trim = 0mm 0mm 0mm 0mm, clip, width=1.0\columnwidth]{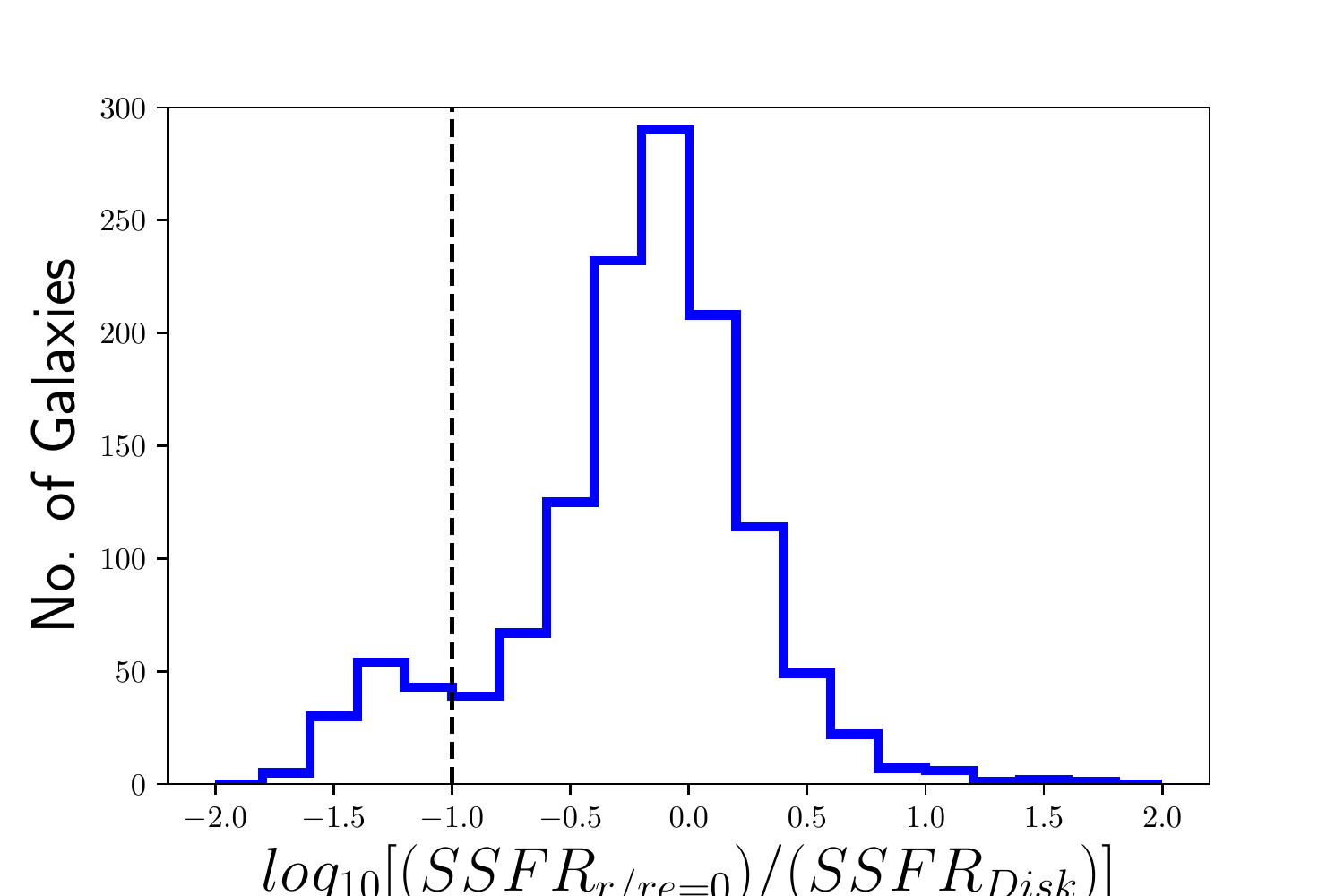}
	\centering
	\caption{\small
		Histogram showing the ratios between the SSFR in the centre most radial bin and the mean SSFR beyond $r/r_e=0.75$. We show with a dashed line the cut between the centrally suppressed and unsuppressed galaxies, which marks where the disk has SSFR is approximately 10 times higher than the core of the galaxy.
		\label{fig:Core_Disk}}
\end{figure}

We show the mean profiles for centrals and satellites in the stellar mass bins in the same panel in Figure \ref{fig:Difference_Mass}, along with the fractional differences between these profiles. The satellite galaxies have lower SSFRs than the centrals in all the stellar mass bins. In the low $M_*$ bin the satellites have $\log_{10}(SSFR) = -10.32 \pm 0.11$ and the centrals $\log_{10}(SSFR) = -10.22 \pm 0.08$. In the medium $M_*$ bin the satellite SSFR is $\log_{10}(SSFR) = -10.49 \pm 0.17$ compared to $\log_{10}(SSFR) = -10.39 \pm 0.14$ for the centrals. There is a large drop in both the satellites and centrals to the high $M_*$ bin, to $\log_{10}(SSFR) = -10.72 \pm 0.21$ and $\log_{10}(SSFR) = -10.68 \pm 0.20$, respectively. In the lowest mass galaxies the satellites have lower SSFR at all radii than the centrals. In the medium mass bin the satellite have lower SSFR at all radii, but in the cores of the galaxies it appears that the satellites are not as suppressed as the centrals. In the highest mass bin, we see that the satellites have higher SSFRs in their cores and lower SSFRs at high radii. However due to the large variance in the profiles caused by the separation of the galaxies which do and do not exhibit central suppression, it is difficult to tell whether the differences seen in the cores of these galaxies are significant. As the central suppression appears to be strongly related to mass, the differences between centrals and satellites could be due to different stellar mass distributions within each bin, however we have checked the distributions and found that this is not the case.

We desire to determine a way to split galaxies between those that have flat profiles or are `Unsuppressed' and those that are `Centrally Suppressed' \footnote{We choose to describe these galaxies as `Centrally Suppressed' as it follows from analysis of integrated galaxy properties. It is common to define some cut in specific star formation rate to divide galaxies into quenched and star forming, we are simply applying similar nomenclature to the local scale.}. In Figure \ref{fig:Core_Disk} we show the ratio between the SSFR in the centre radial bin and the mean SSFR beyond $r/r_e=0.75$ (i.e. in the galaxy disk) for the full galaxy sample. This figure shows that this ratio is bimodal, with most galaxies being evenly distributed around $log_{10}[SSFR_{r/r_e=0} / SSFR_{disk}] = 0$, which represents a flat profile, and a small population of galaxies around $log_{10}[SSFR_{r/r_e=0} / SSFR_{disk}] = -1.25$. We mark on this plot with a dashed line the cut we make between centrally suppressed and unsuppressed galaxies, where the SSFR in the disk is approximately 10 times the SSFR in the centre of the galaxy. We also define galaxies with a central SSFR of $log_{10}(SSFR) < -11.5$ as centrally suppressed, because without this cut the lowest SSFR galaxies in the sample can be classified as unsuppressed.

The higher SSFRs in the centres of high mass satellites could be due to galaxies which have enhanced star formation in their cores, compared to their disks. This would counteract the affect of the centrally suppressed galaxies lowering the mean SSFR, leading to a higher mean SSFR in satellites compared to centrals. We investigate this possibility in Section \ref{Supp_Cen_Sat}.

\subsection{SSFR Profiles at fixed $\sigma_0$}

\begin{figure*}
	\includegraphics[trim = 0mm 0mm 0mm 0mm, clip, width=1.0\textwidth]{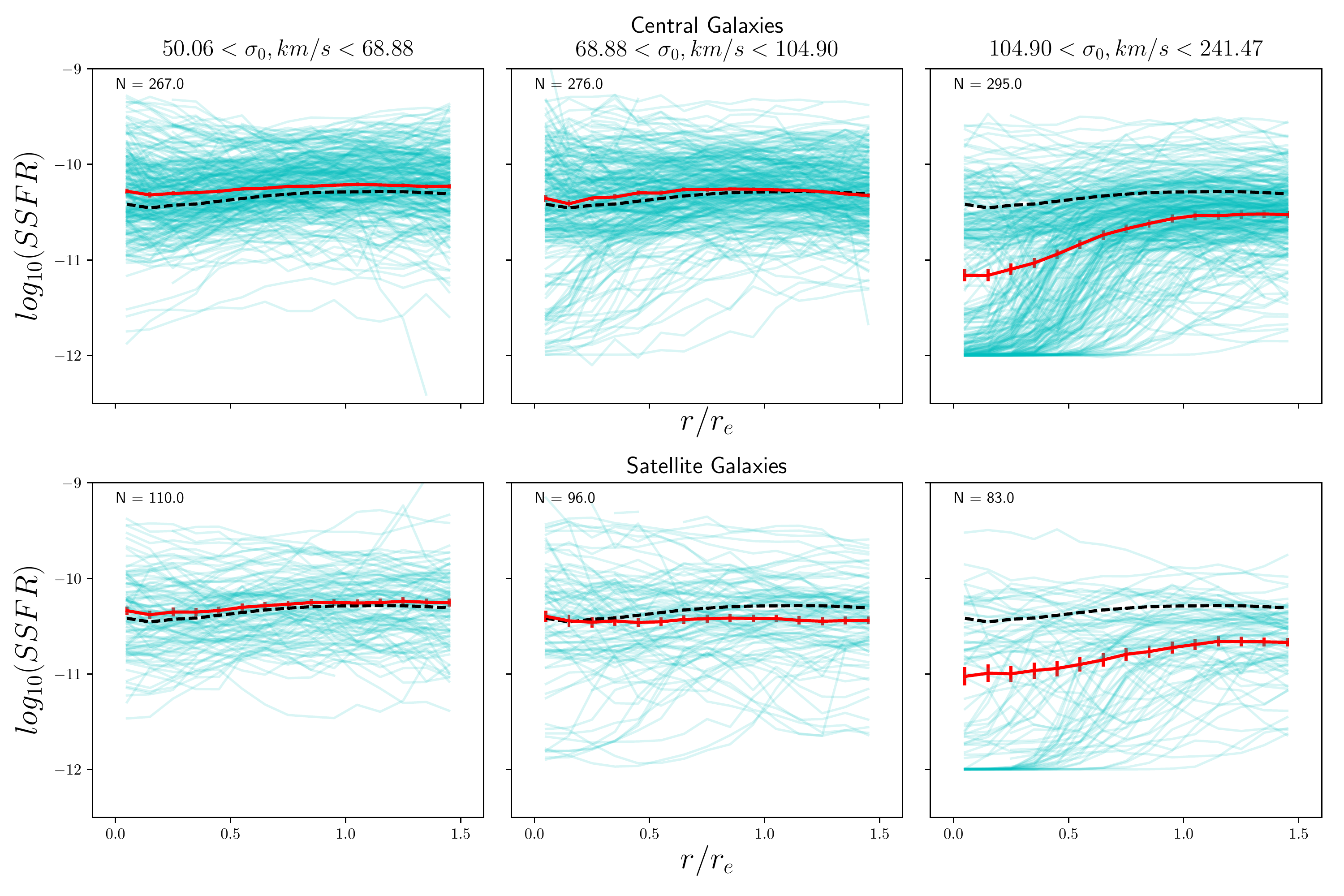}
	\centering
	\caption{\small
		The radial SSFR profiles in three bins of $\sigma_0$. The individual profiles are shown by the cyan lines and the mean profile in the bin is shown by the solid red line. The dashed black line shows the mean profile of all galaxies in the sample. The number of galaxies in each bin is shown in the top left corner of each panel. The top row is the central galaxies and the bottom row is the satellite galaxies. The error bars are calculated from the scatter in 1000 bootstrap resamplings.
		\label{fig:Vo_Profiles}}
\end{figure*}

\begin{figure*}
	\includegraphics[trim = 0mm 0mm 0mm 0mm, clip, width=0.8\textwidth]{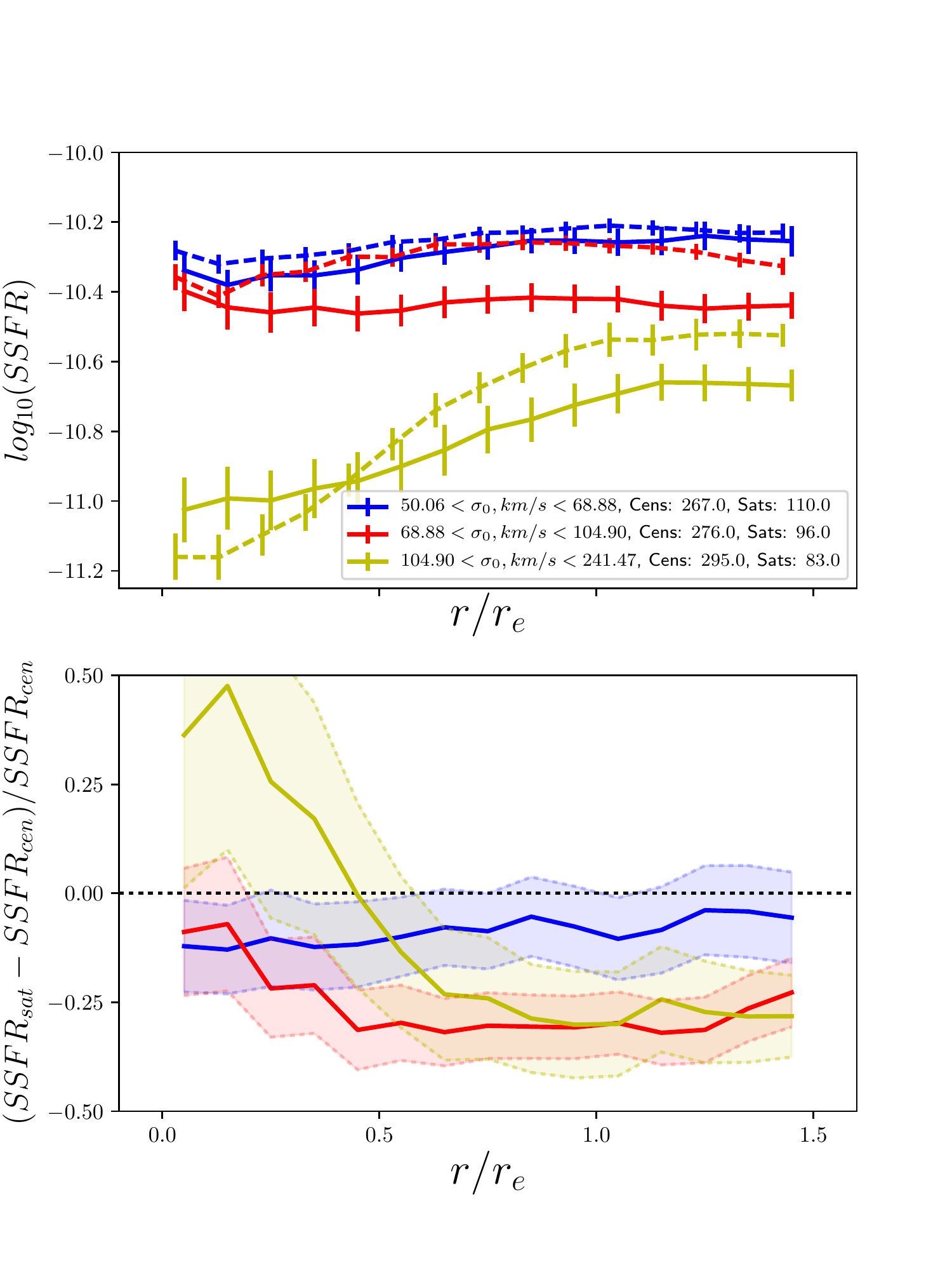}
	\centering
	\caption{\small
		(Top) The mean radial SSFR profiles of central (dashed) and satellite (solid) lines in bins of $\sigma_0$. (Bottom) The fractional difference between the central and satellite mean profiles in bins of $\sigma_0$. The shaded regions and error bars represent the $1-\sigma$ scatter in 1000 bootstrap resamplings.
		\label{fig:Difference_Vo}}
\end{figure*}

In \cite{2017MNRAS.468..333S}, we showed that core velocity dispersion can be a more reliable tracer of environment driven evolution of galaxies than stellar mass. $\sigma_0$ is invariant under environmental processes such as minor mergers and gas stripping, which lead to changes in the mass and size of galaxies. As such we repeat the analysis from the previous section, but instead split galaxies by their core velocity dispersions.

We show the central and satellite profiles in Figure \ref{fig:Vo_Profiles}, using the same plot style as in the previous section. In the lowest $\sigma_0$ bin, we see that the mean profile for centrals and satellites is relatively flat, there are a small number of central galaxies with suppressed cores, but no satellites. In the medium $\sigma_0$ bin the mean profile has a slight downward trend and we once again see an increase in the number of galaxies with suppressed cores, the satellites have a flat profile. In the highest $\sigma_0$ bin there are a large number of centrally quenched galaxies which significantly affect the mean profiles of both centrals and satellites, while the outer profile has remained flat.

We compare the mean profiles and fractional differences between the mean satellite and central profiles in the three $\sigma_0$ bins in Figure \ref{fig:Difference_Vo}. The satellite galaxies generally have lower SSFRs than the centrals. The low $\sigma_0$ bins have similar average SSFRs of $\log_{10}(SSFR) = -10.30 \pm 0.11$ and $\log_{10}(SSFR) = -10.25 \pm 0.08$, for satellites and centrals respectively. In the medium $\sigma_0$ bin the satellite SSFR is 0.1 dex lower, at $\log_{10}(SSFR) = -10.43 \pm 0.16$ for the satellites compared to $\log_{10}(SSFR) = -10.30 \pm 0.12$ for the centrals. There is a large drop in both the satellites and centrals to the high $\sigma_0$ bin, to$\log_{10}(SSFR) = -10.82 \pm 0.24$ and $\log_{10}(SSFR) = -10.76 \pm 0.27$, respectively. It appears that $\sigma_0$ is a better predictor for SSFR than stellar mass, which was also found in \cite{2012ApJ...751L..44W}.

In the low $\sigma_0$ bin, the satellites have $\sim10\%$ less star formation out to $r/r_e=1.5$, where the satellite profiles turn upward slightly and become more star forming than the centrals. In the medium $\sigma_0$ bin, we see that the satellites are less star forming at all radii, however at low radii it appears that the satellites exhibit less core suppression than the centrals as the fractional difference turns towards zero. In the high $\sigma_0$ bin the centrals have higher SSFRs at all radii, except in the cores where the satellites appear to have less suppression, however the scatter in the fractional difference is very high, owing to the large split in SSFRs between galaxies with and without suppressed cores.

\section{Quenching Mechanisms}

\subsection{Centrally Suppressed Galaxies}
\label{CenSupp}

As we have shown in the previous sections, the profile shapes seen in our sample are broadly bimodal. There are galaxies which have flat profiles, and those that have profiles which are centrally suppressed. We have also shown that in the fractional differences between the mean central and satellite SSFR profiles there appears to be two competing effects which are suppressing the star formation in different ways. There is a suppression effect at all radii upon satellite galaxies and some enhancement in the centres of satellites at high mass which may be actual enhancement of star formation or due to less satellites being centrally suppressed. In this section we will explore the populations of centrally suppressed and unsuppressed galaxies separately. 

\begin{figure*}
	\includegraphics[trim = 0mm 0mm 0mm 0mm, clip, width=1.0\textwidth]{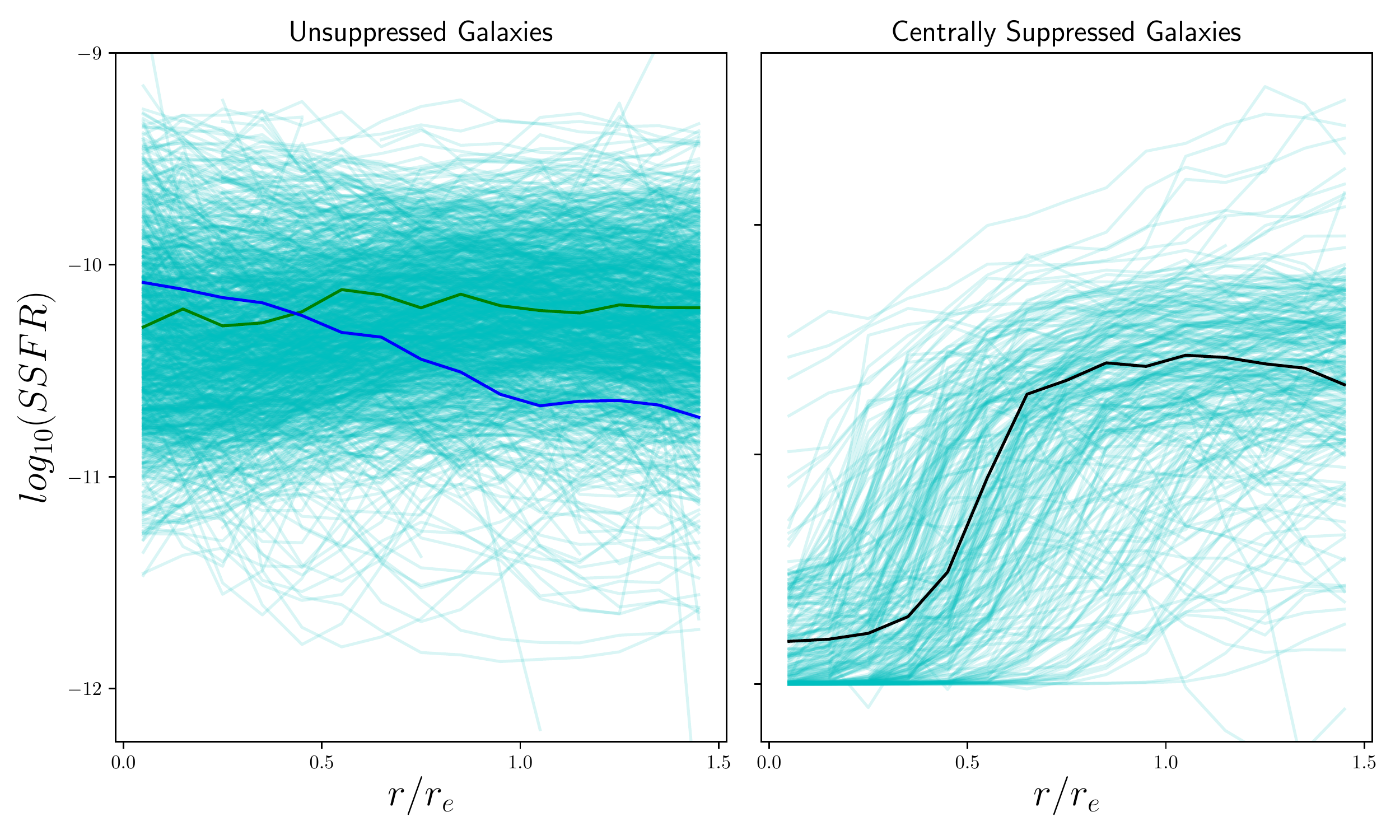}
	\centering
	\caption{\small
		The radial SSFR profiles of galaxies in our sample which are centrally suppressed (left) and unsuppressed (right), as defined using the classification from Figure \ref{fig:Core_Disk}. In the two panels, we highlight 'typical' profiles which fit the Centrally Suppressed (black), Unsuppressed (green) and Enhanced (blue) definitions.
		\label{fig:SplitProfiles}}
\end{figure*}

To demonstrate this split, we plot the radial profiles of the split populations in Figure \ref{fig:SplitProfiles}. The non-suppressed galaxies have predominantly flat profiles, however there is a subpopulation of galaxies which have enhanced SSFR in their cores and a falling profile. The centrally suppressed galaxies appear to be made of two groups, those with linear rising profiles and those which have flat profiles in their outer regions that drop off sharply towards the central bulge. There are also a small number of galaxies which are centrally suppressed by our definition, but in fact exhibit some rejuvenation in their cores.

\begin{figure}
	\includegraphics[trim = 0mm 0mm 0mm 0mm, clip, width=1.0\columnwidth]{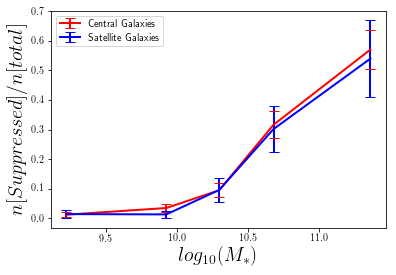}
	\centering
	\caption{\small
		We show the fraction of centrals (red) and satellites (blue) which are centrally suppressed, with respect to Stellar Mass.
		\label{fig:Centrals_Quenched_Properties}}
\end{figure}

In Figure \ref{fig:Centrals_Quenched_Properties} we show the fraction of central and satellite galaxies which are centrally suppressed in bins of stellar mass. We find that there is no difference in the fraction of centrally suppressed galaxies at fixed mass between the central and satellite population. This figure implies then that the mechanisms behind the central suppression are independent from environment completely, and depend only on the galaxy's internal properties. We also see a strong dependence on stellar mass for the fraction of suppressed galaxies, with essentially no galaxies at low mass exhibiting central suppression and ~50\% showing suppression at high masses. This relationship holds when the fractions are instead calculated at fixed $\sigma_0$.

\begin{figure*}
	\includegraphics[trim = 0mm 0mm 0mm 0mm, clip, width=1.0\textwidth]{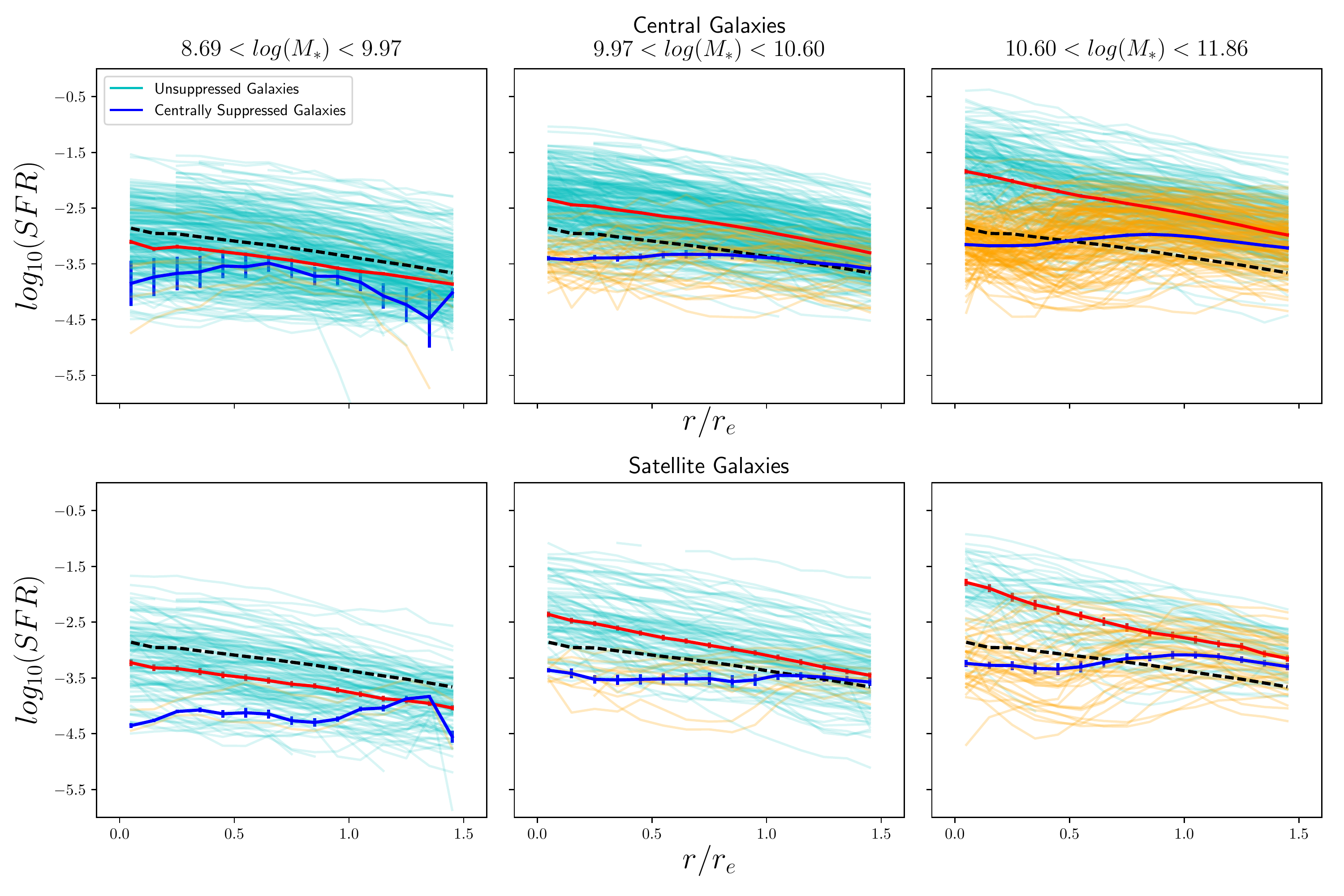}
	\centering
	\caption{\small
		The radial SFR profiles in three bins of stellar mass. We split galaxies based on their core suppression; centrally suppressed galaxies are shown in orange, with the solid blue lines indicating their means; the unsuppressed galaxies are shown in cyan with red lines for their means. The dashed black line shows the mean profile of all galaxies in the sample. The top row is the central galaxies and the bottom row is the satellite galaxies. The error bars are calculated from the scatter in 1000 bootstrap resamplings.
		\label{fig:SFR_Profiles}}
\end{figure*}

One explanation for these centrally suppressed galaxies may be that we are simply tracing the existence of large bulges which formed a long time ago. This would manifest as mass profiles which increase dramatically in the centres of galaxies and SFR profiles which show a simple exponential decrease. When the mass and SFR profiles are combined to produce the SSFR profiles, we would see the characteristic centrally suppressed galaxies. To test whether this is the case we show the SFR profiles for central and satellite galaxies in Figure \ref{fig:SFR_Profiles}. This figure shows the increase in total SFR with stellar mass we demonstrated in \ref{fig:Mass_GroupL_SFR}. We show the unsuppressed and centrally suppressed galaxies with different colour lines in Figure \ref{fig:SFR_Profiles}. There is a clear difference in the SFR profiles of suppressed and unsuppressed galaxies, the centrally suppressed galaxies have lower SFR in their cores than their disks, and have lower SFR than the unsuppressed galaxies at all radii. This Figure shows that the differences in the SSFR profiles are not simply due to differences in mass distribution, but also reflect lower instantaneous star formation. The bimodality is not as strong in SSFR profiles, which is due to the fixed SSFR limit in the $D_n4000$ model, as the centrally suppressed galaxies have a `flat' SSFR in their cores, the increasing mass profile causes the SFR profile to turn upwards, this artefact of the SSFR-$D_n4000$ model masks the centrally suppressed galaxies slightly.

\subsection{Comparison of Centrals and Satellite Profiles}
\label{Supp_Cen_Sat}

With the population split into centrally suppressed galaxies and unsuppressed galaxies, we can revisit the SSFR profiles and determine the quenching effects operating on these different classes of galaxies. By studying the unsuppressed galaxies we can gain a better understanding of the processes which produce the reduction in SSFR at all radii in satellites compared to centrals. Studying the centrally suppressed galaxies we can find if there is a difference in the amount of core suppression which happens in satellites and centrals.

\begin{figure}
	\includegraphics[trim = 0mm 0mm 0mm 0mm, clip, width=1.0\columnwidth]{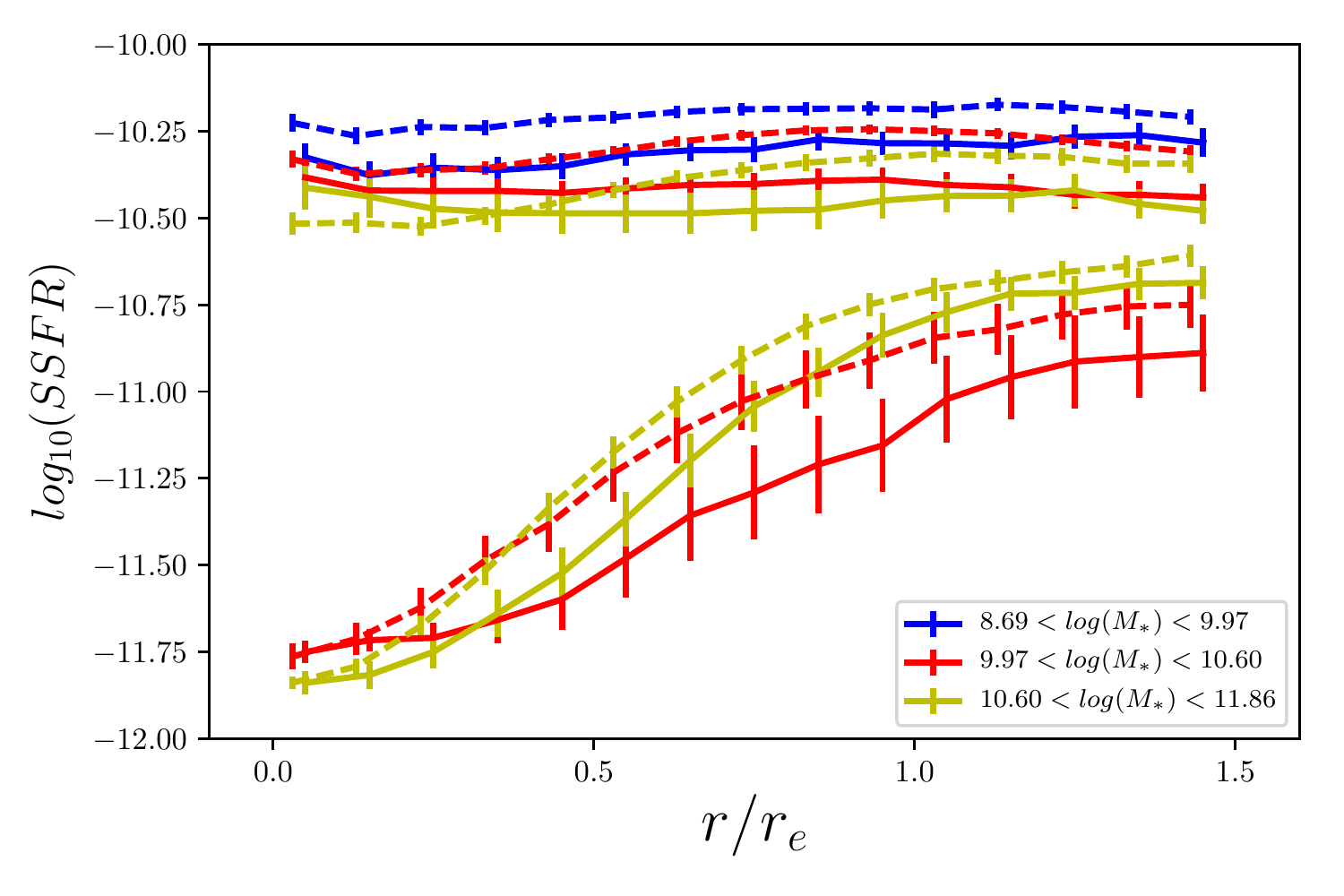}
	\centering
	\caption{\small
		The mean SSFR profiles of centrally suppressed and unsuppressed galaxies. The upper set of lines are the unsuppressed galaxies, while the lower lines are the suppressed galaxies. Satellite profiles use solid lines and centrals use dashing lines. We do not include the low mass bin for the suppressed galaxies. We used the same three stellar mass bins as in Figure \ref{fig:Mass_Profiles}.
		\label{fig:Suppressed_Profiles}}
\end{figure}

In Figure \ref{fig:Suppressed_Profiles} we show the mean profiles of galaxies, split by whether they are centrally suppressed or not. Central galaxies are shown with solid lines and satellites with dashed lines, with the upper set of lines representing the unsuppressed galaxies and the lower lines the suppressed galaxies. We use the same mass binning scheme from Section \ref{Mass}. Note that we do not include the profiles for low mass centrally suppressed galaxies, as there are too few galaxies in this bin to draw reliable conclusions. Firstly we can see that the centrally suppressed galaxies actually have reduce SSFRs at all radii compared to the unsuppressed galaxies, not just in their cores. This is a crucial point, as it suggests that central suppression leads to external suppression, or at least that if fractional growth is low in the centre of galaxies it will be low in the outskirts. The low SSFRs in the outskirts of suppressed galaxies is not a selection effect either, as the ratio we use to divide the sample would certainly allow galaxies with SSFRs 2 or 3 dex higher in their disks, comparable to unsuppressed disks.

For the unsuppressed galaxies the low mass profiles are very similar to the profiles in the low mass bin for the full sample, due to there being very few centrally suppressed galaxies in this bin. The low mass satellites have a very flat profile, which has lower SSFR at all radii than the centrals in this bin, the central profile is also flat. In the medium mass bin the satellites appear to experience suppression at all radii compared to the centrals. In the high mass bin the satellites have higher SSFRs in their cores than the centrals, but beyond $\sim0.5 r_e$ their SSFR is consistently lower. This could be due to high mass satellites that have had some star formation driven into their centres by tidal harassment or some other instability, as it appears that the satellite profile curves upwards, while the central profile curves down.

We see that for the centrally suppressed galaxies, in both the medium and high mass bin the profiles beyond $1.0 r_e$ are quite shallow and rising, and that there is a sharp drop in SSFR towards the centres of the galaxies. The drop appears to happen at a larger radii for the satellite galaxies than the centrals, however both the centrals and satellites approach similar minimum SSFRs, due to the lower limit imposed by our SSFR-$D_n4000$ model. We once again see that there is a suppression of satellite star formation at all radii in the medium and high mass bins.

\begin{figure}
	\includegraphics[trim = 0mm 0mm 0mm 0mm, clip, width=1.0\columnwidth]{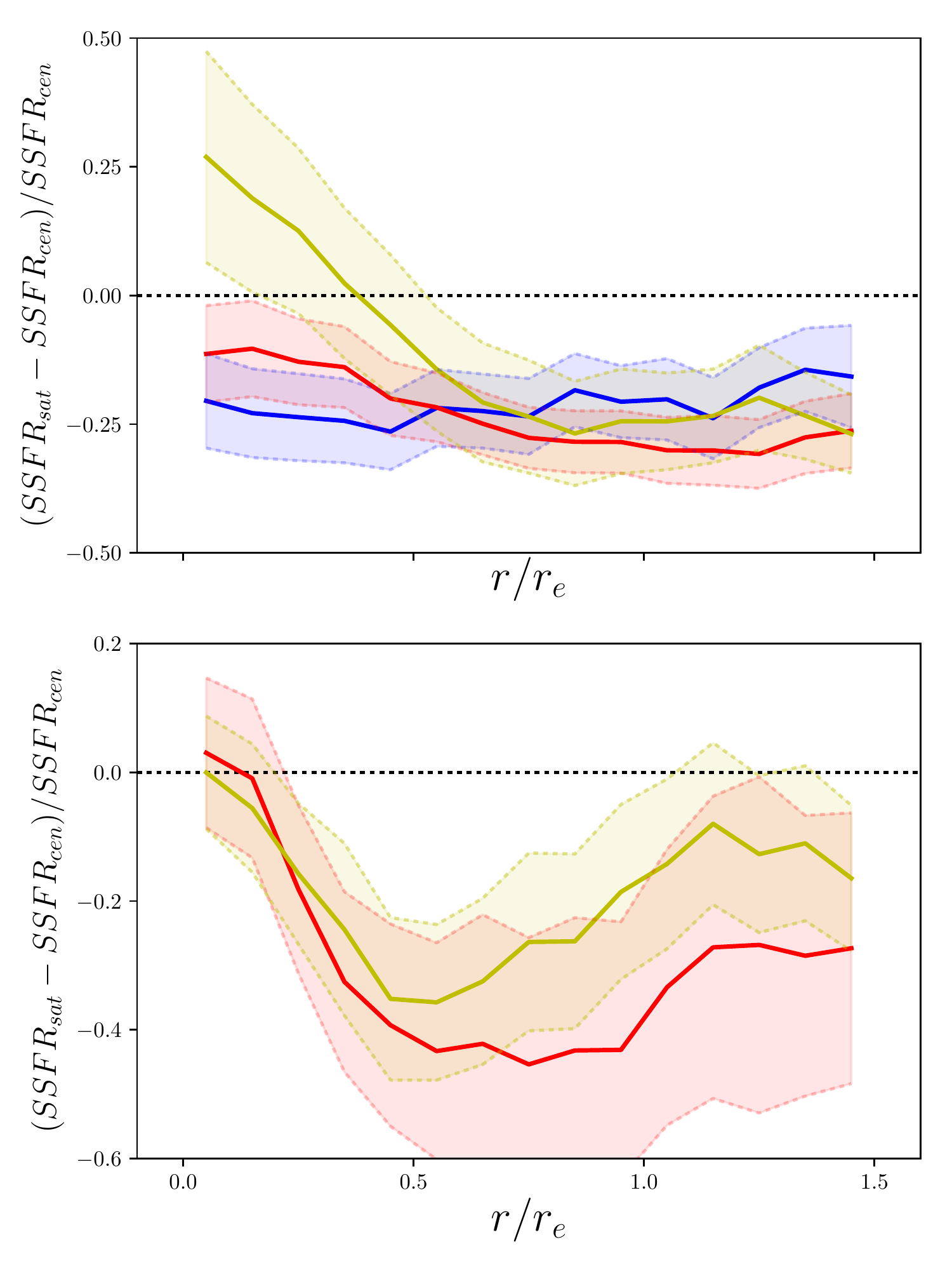}
	\centering
	\caption{\small
		(Top) The fractional differences between central and satellite galaxies in unsuppressed galaxies. We show the $1-\sigma$ scatter from 1000 bootstrap resamplings as the shaded area.
		(Bottom) The fractional differences between central and satellite galaxies in centrally suppressed galaxies. We show the $1-\sigma$ scatter from 1000 bootstrap resamplings as the shaded area.
		\label{fig:Suppressed_Differences}}
\end{figure}

In Figure \ref{fig:Suppressed_Differences} we show the fractional differences between the central and satellite galaxies in bins of mass, split by centrally suppressed and unsuppressed. The unsuppressed galaxies show a roughly uniform decrease in SSFR for satellites compared to centrals, except in the cores of high mass galaxies. For the centrally suppressed galaxies we also see a suppression at all radii in the satellites, though the SSFRs in the cores of the galaxies are approaching parity due to the lower limits of our SSFR-$D_n4000$ model. This uniform suppression of satellites could be a signature of strangulation (\cite{1991PASP..103..390V, 2002AJ....124..777E}), which we discuss further in Section \ref{EnvQuench}.

\begin{figure}
	\includegraphics[trim = 0mm 0mm 0mm 0mm, clip, width=1.0\columnwidth]{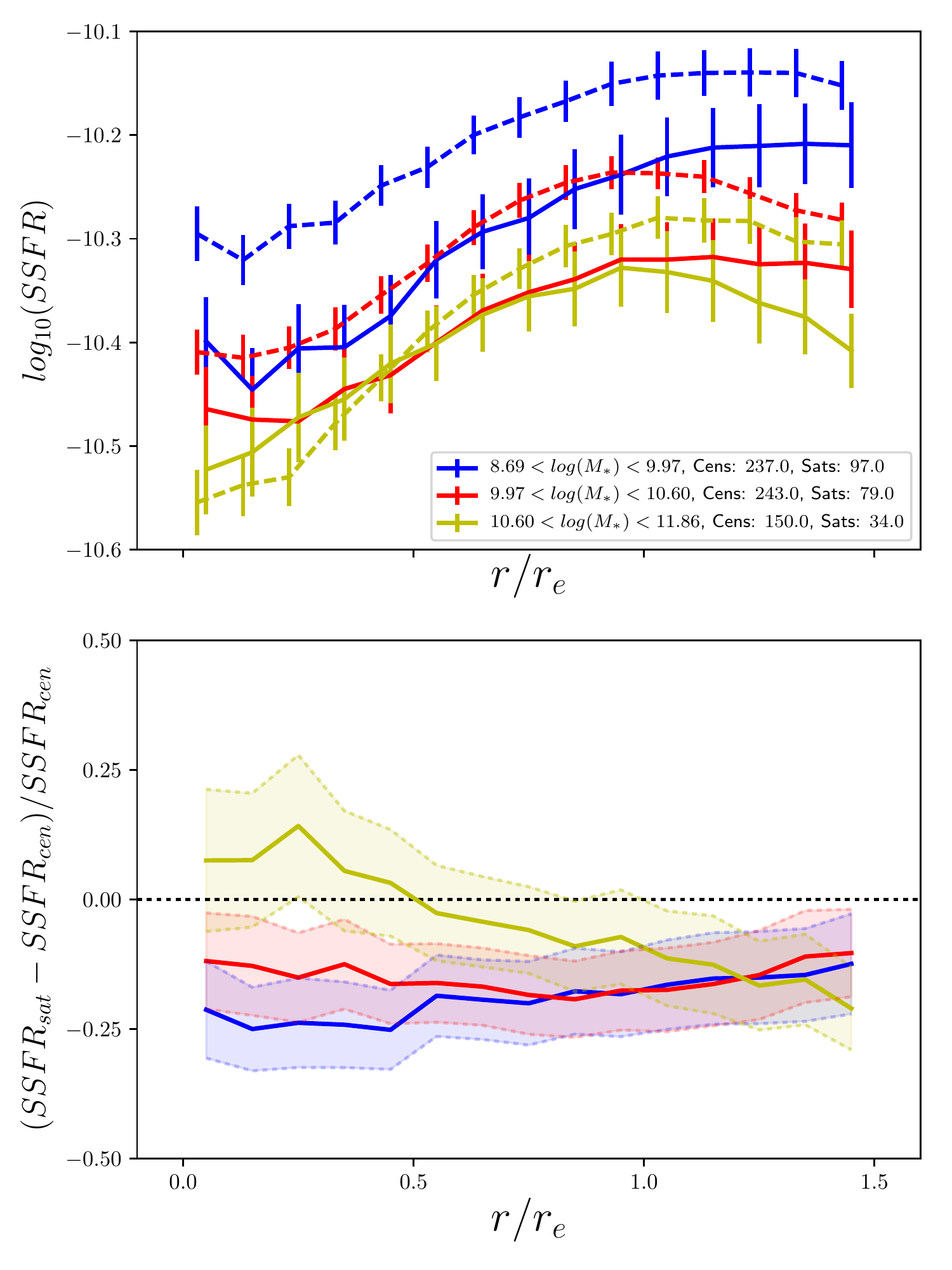}
	\centering
	\caption{\small
		(Top) The mean profiles for unsuppressed galaxies in bins of stellar mass, with the enhanced galaxy population removed. The error bars are calculated from the scatter in 1000 bootstrap resamplings and the stellar mass bins are the same as those from \ref{fig:Mass_Profiles}. Note the different scale in the y-axis compared to Figure \ref{fig:Suppressed_Profiles}
		(Bottom) The fractional differences between central and satellite galaxies in unsuppressed galaxies with centrally enhanced galaxies removed. We show the $1-\sigma$ scatter from 1000 bootstrap resamplings as the shaded area.
		\label{fig:Enh_Differences}}
\end{figure}

Considering the effect that galaxies with enhanced central star formation may have on these results, we devise a additional classification for those galaxies. Profiles where the SSFR in the central radial bin is $0.5$ dex higher than any other radial bin are classified as centrally enhanced. We find that 183 galaxies are centrally enhanced using this classification, they are predominantly star forming galaxies, rather than composite. The fraction of enhanced galaxies decreases with stellar mass and satellites are more likely to be enhanced than centrals. At low mass $18 \pm 8 \%$ of satellites are enhanced, compared to $14 \pm 5 \%$ of centrals, at high mass we find that $14 \pm 3 \%$ of satellites have enhancement and only $6 \pm 1 \%$ of centrals do. We provide the fractional differences between central and satellite profiles of centrally suppressed and unsuppressed galaxies, with those that meet the additional enhanced criteria removed in Figure \ref{fig:Enh_Differences}. The fractional differences for suppressed galaxies remain the same, however for the unsuppressed galaxies we see that the difference in the medium mass bin flattens and that the difference in the central radius bin of the high mass galaxies falls to zero. The exact causes of this enhancement is not clear, neither is the increased fraction in satellite galaxies. We briefly discuss this in Section \ref{EnvQuench}, but would like to note that this will be the subject of further study in a future work.

\subsection{Environmental Quenching}
\label{EnvQuench}

Throughout this paper we have compared the profiles of central and satellite galaxies, as they largely reside in different kinds of environments. At fixed mass, central galaxies are found in lower density environments than satellites, since a satellite of equal mass would require a more massive central to be present in the group. Satellites however are found in denser environments and are acted upon by a number of processes which can shut down star formation, such as ram pressure stripping, tidal stripping and strangulation (\cite{1972ApJ...176....1G, 1999MNRAS.308..947A, 2000ApJ...540..113B, 2002MNRAS.334..673L, 2004MNRAS.353..713K, 2004ApJ...613..851K, 2008MNRAS.383..593M, 2008MNRAS.387...79V, 2008MNRAS.389.1619F, 2011MNRAS.415.1797C, 2015A&A...576A.103B, 2015Natur.521..192P}.

Ram pressure stripping generally causes a decrease in star formation rates at large radii and a central concentration of star formation (\cite{2004ApJ...613..851K, 2011MNRAS.415.1797C}). While we do see more satellites with an enhanced central SSFR compared to centrals, we do not see and increase in suppression with radii as we might expect if ram pressure stripping were important. It could be that due to the cuts we made to effective radii in our sample to ensure good SNR we have excluded the regions of satellites which would be most affected by ram pressure stripping. The increased fraction of centrally enhanced galaxies in the satellite population could be a signal of tidal stripping and disruption, which has been shown to drive gas into the centres of galaxies and cause an increase in circumnuclear star formation \cite{1989Natur.340..687H, 2015MNRAS.448.1107M}.

Strangulation has been shown to be an effective method of quenching galaxies and it is theorised to produce a uniform suppression across a galaxy's radius, as opposed to concentrating star formation in the centre or outskirts (\cite{1980ApJ...237..692L, 1991PASP..103..390V, 2002AJ....124..777E,2008MNRAS.383..593M, 2015Natur.521..192P}). We do see a roughly uniform suppression of star formation in satellite galaxies at all radii for low and medium mass galaxies, especially when we remove the effect of centrally suppressed and enhanced galaxies from the sample, indicating that strangulation may be the dominant satellite quenching mechanism. \cite{2008MNRAS.387...79V} argued that strangulation should be the main process by which satellites quench, as opposed to ram pressure stripping or harassment which occur mainly at high dark matter halo mass. Satellites were found to be redder and more concentrated than centrals, but these differences were independent of halo mass. Similar results were found using data from the EAGLE cosmological simulations (\cite{2015MNRAS.446..521S}) by \cite{2017MNRAS.466.3460V}, who studied the gas accretion rates of simulated galaxies and found that satellites in dense environments are less able to replenish their cold gas than centrals, leading to a shut down of star formation. Finally, \cite{2015Natur.521..192P} studied stellar metallicities and ages from local galaxies and concluded that strangulation, with an average time-scale of 4 billion years, is the dominant mechanism behind galaxy quenching.

\subsection{Morphological Quenching}
\label{MorphQuench}

\begin{figure*}
	\includegraphics[trim = 0mm 0mm 0mm 0mm, clip, width=1.0\textwidth]{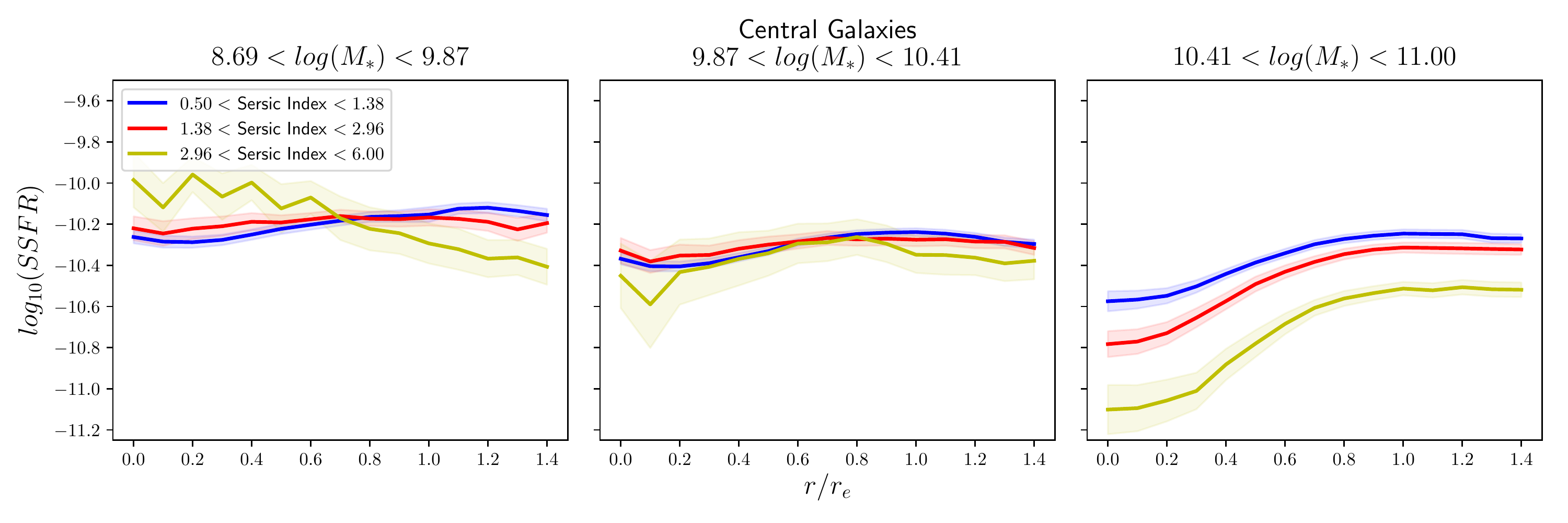}
	\includegraphics[trim = 0mm 0mm 0mm 0mm, clip, width=1.0\textwidth]{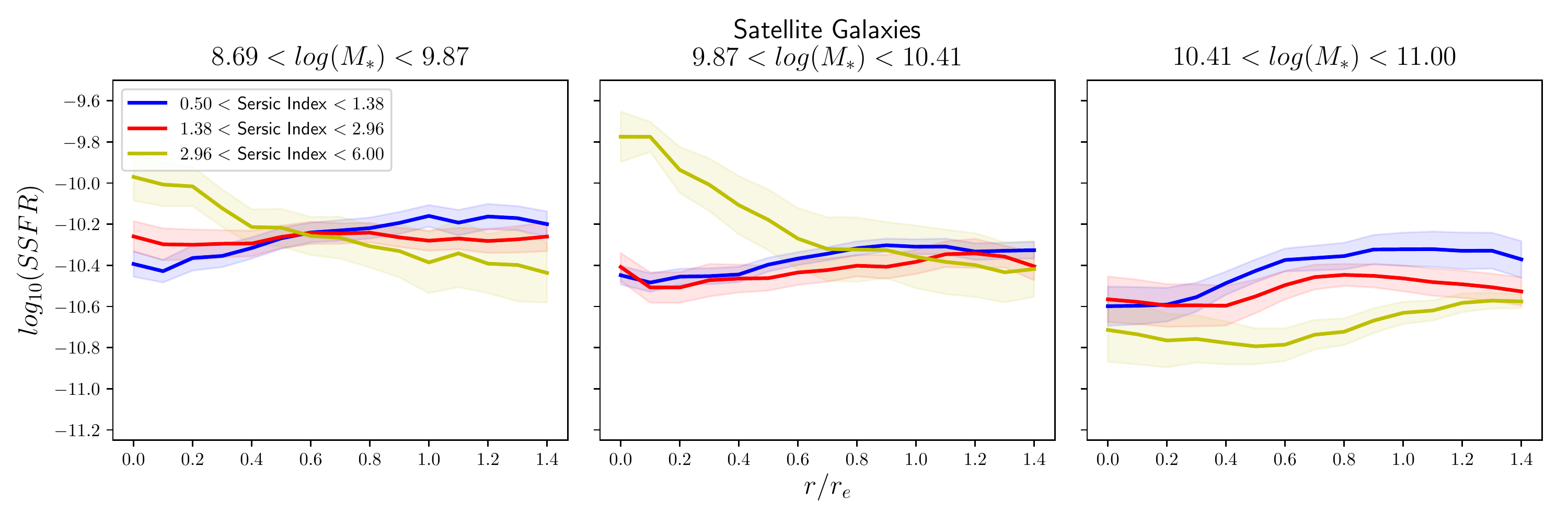}
	\centering
	\caption{\small
		The radial SSFR profiles for central galaxies (top) and satellite galaxies (bottom), in bins of stellar mass and S\'ersic Index. In each bin the blue line represents low S\'ersic index galaxies, red is medium and yellow is high S\'ersic index. The shaded areas represent in the $1-\sigma$ scatter from the mean in 1000 bootstrap resamplings.
		\label{fig:Sersic_Profiles}}
\end{figure*}

\begin{figure*}
	\includegraphics[trim = 0mm 0mm 0mm 0mm, clip, width=1.0\textwidth]{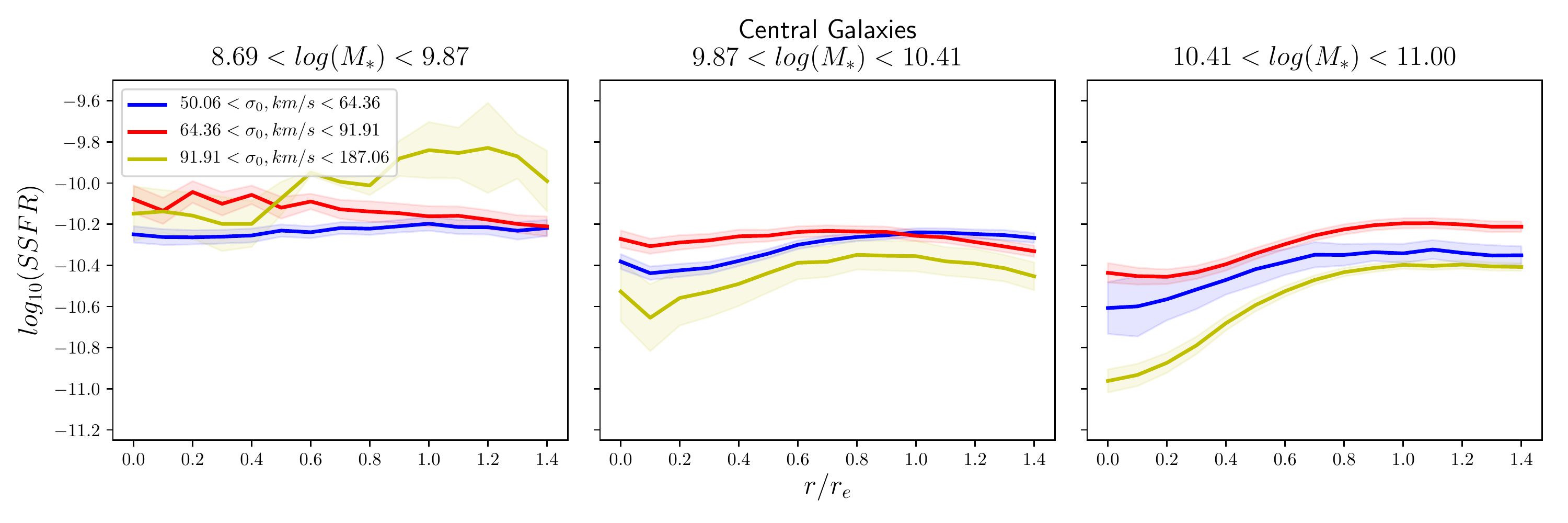}
	\includegraphics[trim = 0mm 0mm 0mm 0mm, clip, width=1.0\textwidth]{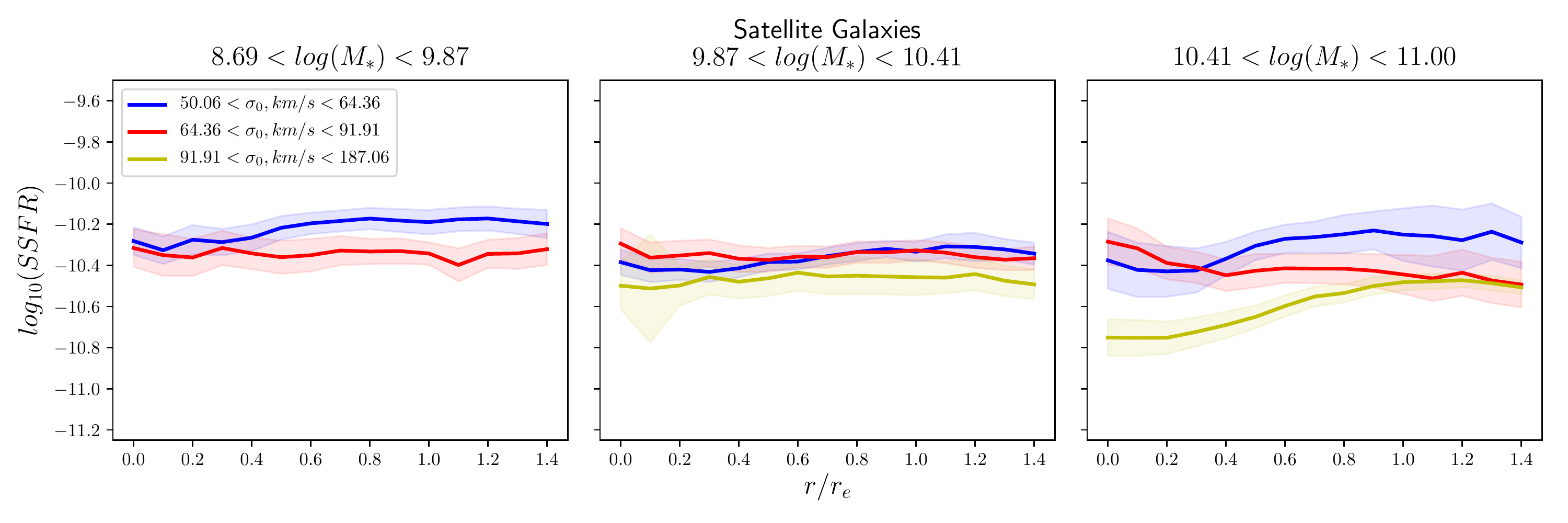}
	\centering
	\caption{\small
		The radial SSFR profiles for central galaxies (top) and satellite galaxies (bottom), in bins of stellar mass and $\sigma_0$. In each bin the blue line represents low $\sigma_0$ galaxies, red is medium and yellow is high $\sigma_0$. The shaded areas represent in the $1-\sigma$ scatter from the mean in 1000 bootstrap resamplings.
		\label{fig:Mass_Sigma_Profiles}}
\end{figure*}

Morphological quenching occurs when a dominant spheroidal component is formed by mergers and other processes, which causes the gas within a galaxy to stabilise against fragmentation and star formation (\cite{2009ApJ...707..250M}). The build up of the bulge then may be what is causing the centrally suppressed galaxies, and may also explain why they have lower star formation rates in their outer regions than non-centrally suppressed galaxies. We now investigate the role of morphology in the suppression of star formation by studying the profiles of galaxies at fixed mass and $r$-band S\'ersic index. If bulge like morphologies do in fact play a role in quenching we would expect to see lower SSFRs at high S\'ersic indices.

In Figure \ref{fig:Sersic_Profiles} we show the mean profiles for central and satellite galaxies in bins of stellar mass and S\'ersic index. The S\'ersic index cuts are such that the lowest bin is mostly pure late type disk galaxies, the medium bin is likely made up of disks with some bulges and bars, while the high S\'ersic index bin is likely dominated by early-type galaxies with large bulges or elliptical morphologies. The shaded areas around the lines represent the $1-\sigma$ scatter from the mean in 1000 bootstrap resamplings.

For the central galaxies in the low and medium mass bins, the low and medium S\'ersic index profiles are very similar, the same goes for the high S\'ersic profile in the medium mass bin. However at low masses the high S\'ersic index profile is quite different, with high SSFR in the centre which falls off towards the edge of the galaxy, as opposed to the flat profiles which appear to be the standard across our sample. In the high mass bin the story is different. While all three profiles are centrally suppressed, we see that the S\'ersic index strongly affects the normalisation of the profile. Higher S\'ersic index galaxies, i.e. those that are more dominated by bulge-like morphologies, have lower SSFRs across their entire profiles.

For the satellite galaxies, many of the properties are the same as the centrals. The low and medium S\'ersic index profiles agree well at low and medium masses, but the medium S\'ersic index galaxies have slightly lower SSFRs at high mass in their disks. The high S\'ersic index satellites have very different profiles compared to the centrals however. We see that the cores of these satellite galaxies are enhanced compared to the general population in both the low and medium mass bins. There also appears to be some enhancement compared to high S\'ersic index centrals in the high mass bin, but not to the same extent as the other profiles. This enhancement may be due to gas being driven into their centres of galaxies by tidal interactions, however it is unclear why this would mainly affect galaxies with high S\'ersic indices.

We also investigate the profiles in bins of stellar mass and $\sigma_0$ simultaneously. We show the mean profiles for central and satellites galaxies in Figure \ref{fig:Mass_Sigma_Profiles}, with galaxies split by mass in the columns and into three bins of $\sigma_0$ in each panel, we omit the low mass-high $\sigma_0$ profile, as there are $< 3$ galaxies in this bin. Velocity dispersion has previously been found to be a better predictor of galaxy colour, bulge mass, bar strength and whether a galaxy is passive or not \cite{2008Ap&SS.317..163D, 2012ApJ...751L..44W, 2016MNRAS.457.2086T, 2017MNRAS.468..333S}. Once again we see that as stellar mass increases the galaxies become more centrally suppressed, in addition we see that in the high mass bins the galaxies with the highest $\sigma_0$ exhibit the strongest suppression of star formation. This suppression occurs both in the cores of these galaxies, but also in the SSFR at all radii. This is particularly strong for central galaxies, where the high mass galaxies with low or medium dispersions are not significantly suppressed compared to the full sample mean and the high dispersion galaxies are very suppressed. One possible explanation for enhancement of high S\'ersic satellites is that it is a selection effect. If these galaxies have very high star formation rates in their centres the light could wash out the disk when the single component fit is attempted, making them seem more bulge dominated.

Combining the results from Figures \ref{fig:Sersic_Profiles} and \ref{fig:Mass_Sigma_Profiles}, we see a strong correlation between the central suppression and bulge dominated morphologies at high and intermediate masses. This suggests that in more bulge like morphologies the galaxies are more likely to be centrally suppressed and that this suppression extends beyond the bulge into the disk. This would appear to agree with the premise of morphological quenching that the large bulge stabilises the gas and prevents star formation. As we do not see an enhancement in the profiles of high $\sigma_0$ satellites, which would be expected if the enhanced galaxies did have very large bulges, this suggests that the S\'ersic index may in fact be skewed higher due to the increased central star formation.

\subsection{AGN Feedback}
\label{AGNfeed}

An alternative to the morphological quenching is the role of AGN feedback. The above results, that galaxies are more likely to be centrally suppressed if they are high mass and have bulge dominated morphologies also imply higher black hole mass, due to the bulge mass-black hole mass relation. The higher mass black holes are more likely to host radio mode AGN which can prevent collapse of gas for star formation and the accretion of gas from the galaxy halo.

\begin{figure}
	\includegraphics[trim = 0mm 0mm 0mm 0mm, clip, width=1.0\columnwidth]{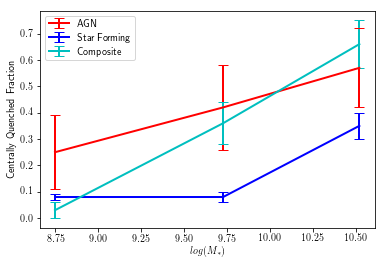}
	\centering
	\caption{\small
		The fraction of galaxies which are centrally quenched, for galaxies which have an integrated BPT classification of AGN, Star Forming and Composite, in three bins of Stellar Mass.
		\label{fig:Quenched_Galaxy_Fractions_AGN}}
\end{figure}

To investigate the role of AGN in core quenching, we can revisit our sample definition and choose to include galaxies which have a BPT classification in their integrated flux as AGN/LINER or low SNR AGN, but which have a total $log_{10}(SSFR) > 10^{-11.5}$. In Figure \ref{fig:Quenched_Galaxy_Fractions_AGN} we show the fraction of galaxies which are centrally quenched in three bins of stellar mass for Star Forming, Composite and AGN galaxies. At all masses, the AGN galaxies are more likely to be centrally suppressed, and in the medium and high mass bins the composites are more likely to be quenched than star forming galaxies as well.

\section{Comparison with previous works}
\label{PrevWorks}

Quenching processes have been widely studied in astronomy, in particular the role of the environments galaxies live in. We will compare our work with some previous studies and draw some conclusions as to what quenching processes may be driving our results.

\cite{2017MNRAS.466.2570B} used the MaNGA survey to reveal what they refer to as eLIER and cLIER galaxies. By studying the emission line properties, they showed that LINER emission is related to old stellar populations, and not necessarily AGN. These galaxy regions do not exhibit star formation, but still emit emission line radiation. cLIER galaxies in particular appear to be late-type spirals which populate the green valley and may be in the process of quenching inside-out. These galaxies are likely related to our centrally suppressed galaxies, which we find to be largely Composite and AGN/LI(N)ER in their BPT classification.

In Belfiore et al. (submitted) we investigated the profiles of specific star formation rate and the equivalent width of Hα of blue cloud and green valley galaxies, with particular emphasis on the properties of cLIER galaxies. We found consistent patterns of central suppression in blue cloud and green valley galaxies, as we have in this work. In addition in Belfiore et al. (submitted) we find that green valley galaxies and cLIER galaxies not only show suppression in their central regions, but are suppressed at all radii, and that this effect is stronger for cLIER galaxies.

The uniform suppression of satellite star formation explains why in \cite{2017MNRAS.465..688G} and \cite{2017MNRAS.465.4572Z} there is no environmental dependence on the gradients of stellar age in MaNGA galaxies, irrespective of whether environment is measured as an environmental density or central/satellite split. In addition, earlier work from \cite{2010MNRAS.404.1775T} showed no dependence on environment for the stellar population properties of early type galaxies, finding that their evolution was driven purely by self-regulation processes related to stellar mass, which is echoed in our findings that the central suppression is independent of environment.

Using the SAMI survey, \citet[][]{2017MNRAS.464..121S}(S17) studied the $H_\alpha$ surface density gradients of 201 star forming galaxies with respect to stellar mass and environmental density. They found that the gradients of $H_\alpha$ surface density steepen as environmental density increases (by a factor of $\sim0.6$ dex in the most massive galaxies).  

\begin{figure*}
	\includegraphics[trim = 0mm 0mm 0mm 0mm, clip, width=1.0\textwidth]{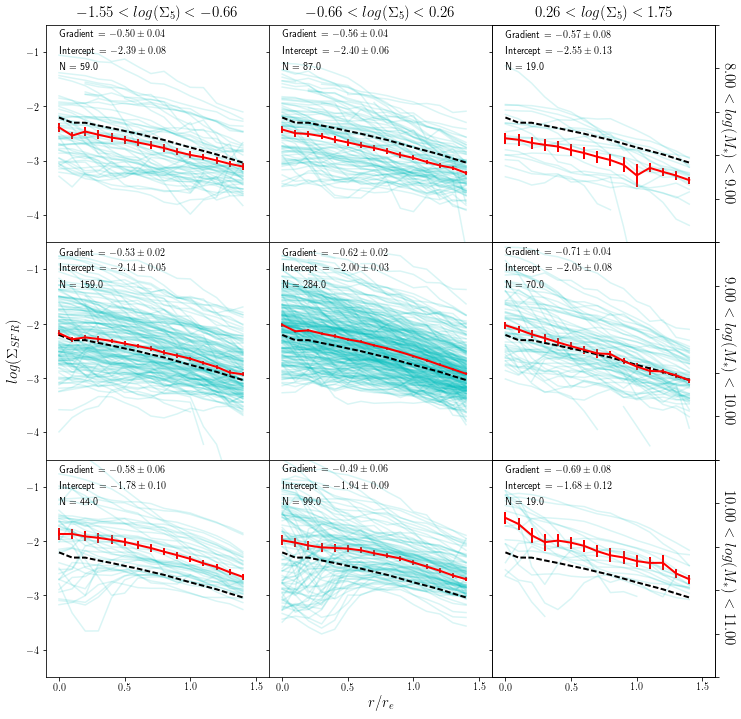}
	\centering
	\caption{\small
		We show the star formation rate surface density in three bins of stellar mass (rows) and three bins of environment density (columns). The cyan lines are the mean radial profiles of the individual galaxies, the solid red line represents the mean profile of all galaxies in the bin and the black dashed line is the mean profile of all galaxies in the sample. We show the properties of a linear fit to the mean profile in the top left corner of each panel.
		\label{fig:Schaefer_Comp}}
\end{figure*}

We provide a direct comparison to S17 in Figure \ref{fig:Schaefer_Comp}, in which we have plotted the profiles of Star Formation Surface Density, $\Sigma_{SFR}$, in bins of stellar mass and nearest neighbour environmental density for star forming galaxies. We use the environmental densities from \cite{2006MNRAS.373..469B}, which are described in Section \ref{OtherCats}. The environment densities in \cite{2006MNRAS.373..469B} were calculated using SDSS galaxies, while S17 uses data from the Galaxies And Mass Assembly (GAMA, \cite{2011MNRAS.413..971D}) survey. The GAMA survey is almost two magnitudes deeper than the SDSS main sample used in \cite{2006MNRAS.373..469B}, meaning that the local density measurements used here at not exactly equivalent. To reconcile this, we have not used the same bins in $\log_{10}(\Sigma_5)$ as S17, but instead we have constructed our bins to contain the same proportion of galaxies in each environment bin as S17. The stellar mass bins were chosen to match those in S17.

We provide the properties of a linear fit to the mean $\Sigma_{SFR}$ profiles and the number of galaxies in each bin in the top corner of each panel, with errors calculated from 1000 bootstrap resamplings. We do see a steepening of the gradients with increasing $\log_{10}(\Sigma_5)$, however this steepening is only significant in the medium mass bin as the gradients in the high and low mass bins are all within $1-2 \sigma$ of each other. We also don't see much central enhancement except in the highest mass and density bin.

Although we have attempted to match the analysis of \cite{2017MNRAS.464..121S} there are a number of differences that may explain our discrepant findings. Of particular importance is the fact that S17 only include spaxels with detectable $H_\alpha$ emission, whereas we include all spaxels, making use of $D_n4000$ where $H_\alpha$ is unavailable. The exclusion of such spaxels in S17 will have the tendency to bias the $\Sigma_{SFR}$ high since these will often by spaxels with low S/N $H_\alpha$ as a result of their low SFRs. This issue most prominently affects the profiles in the central regions of higher mass galaxies where the SFR may be lower if there is a bulge present, our centrally suppressed galaxies. If we also exclude such spaxels then we do indeed see much more central enhancement, typically at high mass and density, which does increase the gradients, although the trend with environment remains present only in the intermediate mass bin. Another difference is that S17 do not take into account possible contamination from AGN/LI(N)ER emission in the individual spaxels in their galaxies (once they have entirely excluded AGN from their sample), which we and \cite{2016MNRAS.461.3111B} have shown is present. Such contamination is again more likely to be present in the central regions of galaxies with a bulge component. Finally as we've already mentioned, we are not using the exact same environmental definition as S17. The higher galaxy density in GAMA means that the fifth nearest neighbour density used by S17 will be probing smaller scales than the measure we have used. It is possible that at these smaller scales a relationship with local overdensity becomes more apparent.

Crucially, it is important to note that the environmental signal of the central/satellite split is much stronger and more significant than the dependence on local environmental overdensity. The relationships between environmental densities and internal properties such as stellar mass and star formation are complex, and two galaxies with similar densities may actually occupy very different conditions and be acted upon by different processes owing to their different locations in the dark matter halos. For example, as we see in Figure \ref{fig:Mass_GroupL_SFR} a satellite and a central occupying the same environmental density can have dramatically different star formation rates, particularly at high densities where the centrals are guaranteed to be very high mass galaxies, whereas the satellites can be very low mass and have very low levels of star formation.

In addition our results show that the profiles of star formation are not linear, with many galaxies exhibiting two or more components in their profiles. In particular our centrally suppressed galaxies would be incredibly poorly fit by a linear profile. S17 argue that star formation becomes more centrally concentrated at higher environment densities

\section{Conclusions}
\label{Conc}

Using IFU data from the SDSSIV-MaNGA survey we have studied the spatial distribution of star formation $1494$ galaxies in the local universe. We have used a two source model to calculate star formation rates using $H_\alpha$ and $D_n4000$, in order to account for emission line contamination in galaxies from AGN and LI(N)ER like sources. The galaxies in our sample were chosen based on their classification in the BPT diagram, using Star Forming and Composite galaxies for the bulk of the work, and introducing a small number of AGN/LI(N)ER galaxies, which passed our total specific star formation rate cut, to study the role of AGN in inside-out quenching.

We have shown that our star formation rate model is internally consistent, by comparing the total star formation rates measured using $H_\alpha$ and $D_n4000$. We have also shown that the total star formation rates agreed well with those calculated for the same galaxies in the MPA/JHU catalog, which use a Bayesian SED fitting method based on $H_\alpha$ and $D_n4000$ from single fibre spectroscopy, aperture corrected to global values using the broadband photometry from SDSS.

Using the radial profiles of specific star formation rate the spatial distribution fo star formation was studied. We binned galaxies based on their internal and external properties and compared the mean profiles in these bins to determine the effect each property had on SSFR. Our main results are as follows:

\begin{itemize}
	\item We found that the SSFR of galaxies decreases with mass and $\sigma_0$. This decrease occurs at both the global scale with total SSFRs, and at the local scale with higher mass and $\sigma_0$ galaxies having lower SSFR at all radii compared to galaxies with low mass and $\sigma_0$.
	\item We revealed the existence of two groups of galaxies, which we have named `Centrally Suppressed' and 'Unsuppressed'. The unsuppressed galaxies have flat profiles in SSFR and can be found at all stellar masses and velocity dispersions. We have defined the centrally suppressed galaxies as having a SSFR in their disk at least 10 times higher than in their core. There is a strong relationship between stellar mass, $\sigma_0$ and whether a galaxy is centrally suppressed or not, with high mass and high $\sigma_0$ galaxies being much more likely to have suppressed SSFR in their cores.
	\item The profiles of the two classes of galaxies showed that the centrally suppressed galaxies actually have suppressed SSFR at all radii, compared to the unsuppressed galaxies. This suggests that central suppression correlates with the suppression of star formation in the outskirts of the galaxy, or at least that low fractional growth in the centre of galaxies means low growth in the outskirts. We find that the mean SSFRs of centrally suppressed galaxies within $0.5 r_e$ of the galaxy centre are $\sim1.25$ dex lower than unsuppressed galaxies, and $\sim0.5$ dex lower beyond $1.0 r_e$. We show that this central suppression is not caused by differences in the mass profiles in these galaxies, as the pattern also emerges in the radial SFR profiles. Centrally suppressed galaxies have lower SFR at all radii compared to unsuppressed galaxies, and have lower SFR in their cores than in their disks.
	\item One possibility is that the suppression is caused by morphological quenching, which we study using the profiles binned by stellar mass and S\'ersic index or $\sigma_0$ simultaneously. These profiles show that both the central suppression and suppression of the disk is strongly correlated to properties which imply large bulges, with high mass-high S\'ersic and high mass-high dispersion galaxies predominantly being centrally suppressed. This result seems to suggest that morphological quenching, where a large bulge component stabilises the gas disk and prevents star formation, may be playing a major role in the lowered SSFRs in the cores and disks of the centrally suppressed galaxies.
	\item We also explored the possibility that suppression of star formation is due to AGN feedback by investigating the fractions of galaxies that were centrally suppressed for star forming galaxies, composites and AGN/LI(N)ER hosts, as characterised by the BPT diagram. We found that at all masses the AGN/LI(N)ER galaxies were more likely to have centrally suppressed SSFRs than star forming galaxies, and that composites were more likely to be suppressed at medium and high masses. AGN feedback also fits in with the increased suppression of core SSFR at high velocity dispersion and S\'ersic index, because the large bulges in these galaxies imply high black hole masses.
	\item Throughout this paper we have compared central and satellite galaxies in order to determine what role environment plays in regulating star formation. We found that central and satellite galaxies are equally likely to have suppressed star formation in their cores, implying that there is no environmental component in that process. However, we did find that satellites do have suppressed SSFRs compared to central galaxies at all radii. This lowered star formation in satellite galaxies is most likely caused by strangulation, which has previously been found to be a likely candidate for satellite quenching. We do not see any suppression in the outskirts of satellites that would be related to ram pressure stripping. We do find that there are a population of galaxies with enhanced star formation in their centres which are more likely to be satellites than centrals, this may be due to tidal harassment driving gas into the centres of satellites, however this is currently unclear and will be the subject of a future study.
	\item Finally, we compared our work to that of \cite{2017MNRAS.464..121S}, who found a steepening of the SFR surface density gradients with 5-th nearest neighbour environment density. We too see a small amount steepening, however we find that it is not statistically significant.
\end{itemize}

Our results in this work show the power of IFU surveys in analysing the spatial properties of galaxies for studying the mechanisms behind the shut down of star formation. We have found evidence of inside-out quenching driven associated with AGN/LI(N)ER like emission, implying suppression of star formation via AGN feedback. In addition we have observed a uniform suppression of star formation in satellite galaxies, indicative of strangulation of cool gas supplies.

\section*{Acknowledgements}

Funding for the Sloan Digital Sky Survey IV has been provided by the Alfred P. Sloan Foundation, the U.S. Department of Energy Office of Science, and the Participating Institutions. SDSS acknowledges support and resources from the Center for High-Performance Computing at the University of Utah. The SDSS web site is www.sdss.org.

SDSS is managed by the Astrophysical Research Consortium for the Participating Institutions of the SDSS Collaboration including the Brazilian Participation Group, the Carnegie Institution for Science, Carnegie Mellon University, the Chilean Participation Group, the French Participation Group, Harvard-Smithsonian Center for Astrophysics, Instituto de Astrof\'isica de Canarias, The Johns Hopkins University, Kavli Institute for the Physics and Mathematics of the Universe (IPMU) / University of Tokyo, Lawrence Berkeley National Laboratory, Leibniz Institut f\"{u}r Astrophysik Potsdam (AIP), Max-Planck-Institut f\"{u}r Astronomie (MPIA Heidelberg), Max-Planck-Institut f\"{u}r Astrophysik (MPA Garching), Max-Planck-Institut für Extraterrestrische Physik (MPE), National Astronomical Observatories of China, New Mexico State University, New York University, University of Notre Dame, Observat\'orio Nacional / MCTI, The Ohio State University, Pennsylvania State University, Shanghai Astronomical Observatory, United Kingdom Participation Group, Universidad Nacional Autónoma de México, University of Arizona, University of Colorado Boulder, University of Oxford, University of Portsmouth, University of Utah, University of Virginia, University of Washington, University of Wisconsin, Vanderbilt University, and Yale University.

M.B. acknowledges funding from NSF/AST-1517006.

VW acknowledges support of the European Research Council from the starting grant SEDmorph (P.I. V. Wild).

\begin{table*}
	\centering
	\caption{The percentage of galaxies which are classified as centrally suppressed in three bins of stellar mass and velocity dispersion, for all galaxies in the sample, central galaxies and satellite galaxies. We include the number of galaxies in each bin in parentheses.}
	\label{tab:EnvPerc}
	\begin{tabular}{llccc}
		&                                                     & \multicolumn{1}{l}{\% of All Galaxies} & \multicolumn{1}{l}{\% of Central Galaxies} & \multicolumn{1}{l}{\% of Satellite Galaxies} \\ \hline
		\multirow{3}{*}{Stellar Mass}                                        			 & $8.09 < log(M_*) < 9.36$                            & $2 \pm 1$ $(19)$                               & $1 \pm 1$ $(9)$                                  & $2 \pm 2$ $(10)$                                   \\
		& $9.36 < log(M_*) < 9.99$                            & $10 \pm 2$ $(75)$                            & $10 \pm 2$ $(55)$                                & $9 \pm 3$ $(20)$                                   \\
		& $9.99 < log(M_*) < 10.99$                          & $39 \pm 4$ $(251)$                            & $41 \pm 5$ $(198)$                                & $37 \pm 7$ $(53)$                                  \\ \hline
		\multirow{3}{*}{Core Velocity Dispersion} 			 			 & $50.6 < \sigma_0, km/s < 68.88$    	     & $2 \pm 1$ $(17)$                              & $2 \pm 1$ $(8)$                                  & $1 \pm 1$ $(9)$                                   \\
		& $68.88 < \sigma_0, km/s < 104.90$  	     & $8 \pm 1$ $(53)$                              & $8 \pm 2$ $(39)$                                  & $8 \pm 3$ $(14)$                                   \\
		& $104.90 < \sigma_0, km/s < 241.47$ 	     & $47 \pm 5$ $(268)$                            & $50 \pm 6$ $(208)$                                & $42 \pm 9$ $(60)$                                 
	\end{tabular}
\end{table*}

\begin{table*}
	\centering
	\caption{The percentage of galaxies which are classified as centrally suppressed in three bins of stellar mass and velocity dispersion, for star forming galaxies, composite galaxies and AGN/LI(N)ER galaxies. We include the number of galaxies in each bin in parentheses.}
	\label{tab:AGNPerc}
	\begin{tabular}{llccc}
		&                                                     & \multicolumn{1}{l}{\% of SF Galaxies} & \multicolumn{1}{l}{\% of Composite Galaxies} & \multicolumn{1}{l}{\% of AGN/LI(N)ER Galaxies} \\ \hline
		\multirow{3}{*}{Stellar Mass}                                        & $8.09 < log(M_*) < 9.36$                            & $1 \pm 1$ $(6)$                               & $2 \pm 2$ $(1)$                                   & $19 \pm 15$ $(3)$                                    \\
		& $9.36 < log(M_*) < 9.99$                            & $4 \pm 1$ $(18)$                               & $35 \pm 7$ $(35)$                                 & $47 \pm 16$ $(14)$                                    \\
		& $9.99 < log(M_*) < 10.99$                          & $25 \pm 4$ $(61)$                             & $57 \pm 7$ $(141)$                                 & $69 \pm 17$ $(41)$                                   \\ \hline
		\multirow{3}{*}{Core Velocity Dispersion} 			 & $50.6 < \sigma_0, km/s < 68.88$    	     & $1 \pm 1$ $(5)$                               & $7 \pm 5$ $(2)$                                   & $15 \pm 15$ $(1)$                                    \\
		& $68.88 < \sigma_0, km/s < 104.90$  	     & $5 \pm 1$ $(23)$                               & $21 \pm 5$ $(19)$                                 & $32 \pm 15$ $(7)$                                    \\
		& $104.90 < \sigma_0, km/s < 241.47$ 	     & $27 \pm 6$ $(56)$                             & $59 \pm 7$ $(156)$                                 & $69 \pm 15$ $(47)$                                  
	\end{tabular}
\end{table*}

\section{Appendix A}

In Tables \ref{tab:EnvPerc} and \ref{tab:AGNPerc} we provide summaries of the percentage of galaxies which are classified as centrally suppressed. Table \ref{tab:EnvPerc} shows the percentages for galaxies in different environments and Table \ref{tab:AGNPerc} shows percentages for different BPT classifications of galaxies.




\bibliographystyle{mnras}
\bibliography{MaNGA_SFR_Bib} 






\bsp	
\label{lastpage}
\end{document}